\documentclass[acmsmall]{acmart}

\usepackage{pdflscape}
\usepackage{multirow}
\usepackage[most]{tcolorbox}
\usepackage{xcolor,soul}
\usepackage{tikz}
\usepackage{orcidlink}
\usepackage{subcaption}
\usepackage{enumitem}
\usepackage{bm}
\usepackage{multirow,pifont,makecell,booktabs,threeparttable,adjustbox,lscape,arydshln}
\usepackage{threeparttable}
\usetikzlibrary{positioning, shapes.geometric}

\newcommand{\enc}[1]{\llbracket {#1} \rrbracket}

\AtBeginDocument{%
  \providecommand\BibTeX{{%
    \normalfont B\kern-0.5em{\scshape i\kern-0.25em b}\kern-0.8em\TeX}}}

\newtcolorbox{mybox}[2][]{%
  sidebyside align = top seam,
  colback      = black!5!white,
  colframe     = black!75!black,
  colbacktitle = gray!85!white,
  title        = #2,#1,
  enhanced,
}

\setcopyright{acmcopyright}
\copyrightyear{2018}
\acmYear{20XX}
\acmDOI{XXXXXXX.XXXXXXX}

\acmJournal{CSUR}
\acmVolume{37}
\acmNumber{4}
\acmArticle{111}
\acmMonth{8}

\begin{document}

\title{A Survey of Secure Computation Using Trusted Execution Environments}


\author{Xiaoguo Li \orcidlink{0000-0001-6496-7454}}
\affiliation{%
  \institution{Singapore Management University}
  \streetaddress{81 Victoria St.}
  \city{Singapore}
  \country{Singapore}
  \postcode{188065}
}
\email{xiaoguoli@smu.edu.sg}

\author{Bowen Zhao \orcidlink{0000-0001-9864-9729}}
\affiliation{%
  \institution{Xidian University}
  \city{Guangzhou}
  \country{China}
  \postcode{510555}
}
\email{zhaobw29@163.com}

\author{Guomin Yang \orcidlink{0000-0002-4949-7738}}
\affiliation{%
  \institution{Singapore Management University}
  \city{Singapore}
  \country{Singapore}
  \postcode{188065}
}
\email{gmyang@smu.edu.sg}

\author{Tao Xiang \orcidlink{0000-0002-9439-4623}}
\affiliation{%
  \institution{Chongqing University}
  \city{Chongqing}
  \country{China}
  \postcode{400044}
}
\email{txiang@cqu.edu.cn}

\author{Jian Weng \orcidlink{0000-0002-7933-9941}}
\affiliation{%
  \institution{Jinan University}
  \city{Guangzhou}
  \country{China}
  \postcode{510632}
}
\email{cryptjweng@gmail.com}

\author{Robert H. Deng \orcidlink{0000-0003-3491-8146}}
\affiliation{%
  \institution{Singapore Management University}
  \city{Singapore}
  \country{Singapore}
  \postcode{188065}
}
\email{robertdeng@smu.edu.sg}

\renewcommand{\shortauthors}{Li et al.}

\begin{abstract}
As an essential technology underpinning trusted computing, the trusted execution environment (TEE) allows one to launch computation tasks on both on- and off-premises data while assuring confidentiality and integrity. This article provides a systematic review and comparison of TEE-based secure computation protocols. 
We first propose a taxonomy that classifies secure computation protocols into three major categories, namely secure outsourced computation, secure distributed computation and secure multi-party computation. To enable a fair comparison of these protocols, we also present comprehensive assessment criteria with respect to four aspects: setting, methodology, security and performance. Based on these criteria, we review, discuss and compare the state-of-the-art TEE-based secure computation protocols for both general-purpose computation functions and special-purpose ones, such as  privacy-preserving machine learning and encrypted database queries. To the best of our knowledge, this article is the first survey to review TEE-based secure computation protocols and the comprehensive comparison can serve as a guideline for selecting suitable protocols for deployment in practice. Finally, we also discuss several future research directions and challenges.
\end{abstract}



\maketitle



\section{Introduction}
The popularity of data-driven analytics has brought exponential data volume growth, especially in the Internet of Things (IoT) era. There will be approximately 572 zettabytes of data produced by 2030 \cite{datagrow}, which is about 30 times that of today. Accompanying the rapid data growth is the continuously evolving computing technologies for performing complex analytical tasks over huge-scale data. To eliminate the burden and cost of deploying and maintaining a high-performance computing infrastructure, many organizations resort to cloud services nowadays.
Nonetheless, under such an off-premises computing model, a cloud server may extract valuable or sensitive information for its own benefit, which could damage the interests of the data owners. Moreover, in computation tasks involving multiple data sources, protecting the interest of each data owner is also an essential requirement. As a general terminology, secure computation refers to technologies that can ensure data privacy and/or integrity throughout the whole computation.  

In general, the existing secure computation frameworks mainly target evaluating a function over encrypted or obfuscated inputs. Its actual form depends on the computation task and setting.  For example, in a multi-party setting, secure computation allows multiple parties with private inputs to evaluate a joint function based on all private inputs, such as a statistical function in a voting scenario or a sorting function in an auction scenario. During the computation, it is necessary to ensure the privacy of each individual input as well as the correctness/integrity of the output. In the outsourced computation setting, secure computation allows a client (or clients) to outsource computation tasks, which could involve sensitive data, to an untrusted cloud server.  
The main challenge of secure computation is: 
{\itshape How can we securely and efficiently evaluate functions on private inputs in the presence of untrusted or malicious parties participating in the computation?}

\noindent\textbf{Cryptographic Approaches for Secure Computation.} In the past four decades, many cryptographic primitives and protocols have been proposed to address secure computation problems. Following Yao's garbled circuits (GC, \cite{yao1986generate}) and GMW's scheme \cite{goldreich1987play} since the 1980s, enormous efforts have been put to improve security and efficiency in secure multi-party computation.
For example, the cut-and-choose mechanism \cite{lindell2007efficient} extended Yao's protocol against malicious adversaries, and dual execution \cite{kolesnikov2013improved} improved GMW's scheme to achieve stronger security. 
Other examples of cryptographic primitives for secure computation include Oblivious RAM (ORAM, \cite{goldreich1996software,stefanov2018path}), homomorphic encryption (HE, \cite{gentry2009fully}), and secret sharing (SS, \cite{shamir1979share}). ORAM-based approaches \cite{wang2014scoram, liu2014automating, doerner2017scaling} build shared ORAM, and then data privacy is achieved by multiple rounds of interactions between the CPU and the shared memory. 
Since all these interactions are encrypted and the shared ORAM re-encrypts and shuffles the data whenever it is accessed, low efficiency becomes a major drawback. HE-based solutions \cite{damgaard2012multiparty, damgaard2013practical} allow computation on ciphertexts directly, however some of them (PHE \cite{paillier1999public}, SWHE \cite{damgaard2012multiparty, damgaard2013practical}) cannot support arbitrary computation, whereas others (e.g., FHE \cite{gentry2009fully}) are still impractical because of the extensive computation cost.
SS-based solutions \cite{cramer2000general, halpern2004rational} split a secret into multiple shares and then distribute them among two or more players. With the shares, one can reconstruct the original secret from them (e.g., by Lagrangian interpolation). 
In recent years, secret sharing has shown its superiority and serves as an essential component in several mixed frameworks for secure computation, such as ABY \cite{demmler2015aby} and ABY2.0 \cite{patra2021aby2}.

Yet, the aforementioned cryptographic approaches are still not widely adopted due to their strong trust assumptions or extensive computation and communication overhead, especially when dealing with complex computation tasks or large datasets. Many approaches (e.g., \cite{cramer2000general,damgaard2012multiparty,veugen2015secure,patra2021aby2,zhao2022soci}) assume the involved parties are semi-honest, meaning those parties, despite being curious about others' private inputs, would follow the protocol steps honestly. 
Such an assumption does not hold in many application domains where parties have incentive to gain more benefits or advantages over their competitors by deviating from the protocol specification.
On the other hand, defending malicious parties using pure crypto-based solutions introduces a large overhead. We note that secure computation protocols that work in the semi-honest setting can be converted to defend malicious participants (e.g., \cite{lindell2007efficient,kreuter2012billion,benhamouda2015implicit}), but with extensive computation and communication costs.

\noindent\textbf{TEE-based Secure Computation.}
Today's growing awareness of Trusted Execution Environment (TEE \cite{platform2013global}) has attracted many interests to explore alternative ways to build practical general/special-purpose secure computation protocols. Thanks to the trusted hardware,  TEE-based secure computation protocols are shown to be more efficient than the traditional crypto-based approaches while offering both confidentiality and integrity guarantees. {\itshape Therefore, TEE serves as a promising technology to enable practical and scalable secure computation for real applications. This article reviews and compares the state-of-the-art TEE-based secure computation protocols and discusses the remaining challenges, open problems and future research directions.}

\subsection{Background of Trusted Execution Environment} \label{subs:tee_background}

TEE provides an isolated environment (called secure {\itshape enclave} \footnote{In the rest of this survey, we use the terms `TEE' and `enclave' interchangeably}) that safeguards processed data by encrypting the incoming and outgoing data. Furthermore, TEE provides mechanisms to ensure the computation is correctly executed with an integrity guarantee. More importantly, TEE protects the data and computation against any potentially malicious entity residing in the system  (including the kernel, hypervisor, etc.). 
Thus, TEE-based secure computation has attracted numerous researchers from academia and industrial communities in the last two decades \cite{lie2000architectural, champagne2010scalable, anati2013innovative, mckeen2013innovative, maas2013phantom, costan2016sanctum} because of three appealing features.
\begin{itemize}
    \item {\itshape General purpose computation.} The supported functions executed in TEE are usually denoted as user-defined programs; thus, it permits a broad spectrum of functions.
    \item {\itshape Integrity and privacy.} TEE provides guarantees of integrity as well as confidentiality. Confidentiality means the attackers cannot obtain sensitive information from the computation, and integrity ensures that the program executes correctly.
    \item {\itshape High performance.} TEE computes the user-defined programs by decrypting all encrypted inputs in enclaves and executing the program on plaintext. Therefore, it enjoys the CPU speed performance, except the overhead for handling the encrypted inputs. 
\end{itemize}

According to the design, current TEE can be classified into two categories: (1) TEE without a secure counter just supports simple stateless functions (e.g., smart cards \cite{NISTsmartcard,poulsen2001directv})  and (2) TEE with a secure counter supports complex stateful functions. In recent years, many TEE designs focus on the latter such as the Trusted Platform Modules (TPM/vTPM) \cite{sumrall2003trusted}, Intel TXT \cite{futral2013fundamental}, Intel SGX \cite{costan2016intel}, ARM’s TrustZone \cite{arm2009arm}, Sanctum \cite{costan2016sanctum}, KeyStone \cite{lee2020keystone} and AMD SEV \cite{kaplan2016amd}. These designs vary significantly in terms of architectural choices, instruction sets, implementation details, cryptographic suites, as well as security features \cite{pass2017formal}. 

\subsubsection{Security Mechanisms}
TEE exhibits a broad range of security features \cite{fei2021security} to achieve confidentiality and integrity of the computation. With these features, TEE has been adopted widely by complex secure systems \cite{choi2019secure}. We present several essential features related to secure computation below.
\begin{itemize}
    \item {\itshape Secure boot \cite{singh2021enclaves}.} When the host starts, secure boot loads immutable and verified images into an enclave to ensure a chain of trust among the enclave images, operating system components, and configurations. Before the operating system starts, a secure boot ensures the TEE is loaded correctly and no hosts, even the hypervisor, can tamper with it.
    \item {\itshape Attestation \cite{anati2013innovative}.}  Attestation establishes trust between an enclave and another enclave or a remote machine  by building a secure channel. Generally, there are two attestation mechanisms: local attestation and remote attestation. While two enclaves on the same platform can attest to each other using local attestation, remote attestation allows a client to authenticate that its program is configured and executed correctly (i.e., not modified or tampered with) in a remote server. 
    \item {\itshape Isolated execution \cite{mckeen2013innovative}.} TEE loads user-defined programs into enclaves and then performs isolated execution for each program. During execution, sensitive data (including code and data) is maintained in EPCs (Enclave Page Caches) within the PRM (Processor Reserved Memory), which is pre-configured in DRAM in bootstrapping phase. The code loaded inside an enclave can access both non-PRM and PRM memory. On the other hand, external programs cannot access any data in the PRM memory.
    \item {\itshape Sealing \cite{anati2013innovative}.} Sealing is a mechanism to let a TEE securely persist and retrieve ``secrets'' on the local disk. Sealing binds the secrets to a specific enclave identity by a sealing key derived from the CPU hardware. Using the sealing key, TEE encrypts the enclave's secrets and stores them on a local disk. Conversely, TEE retrieves the secrets from the disk by the unsealing mechanism.
\end{itemize}

\subsubsection{Vulnerabilities}
Despite its appealing features, TEE is also vulnerable to several attacks, and the survey in \cite{fei2021security} discusses these attacks in details.  
Below we briefly review two major attacks that would affect secure computation.
First, TEE may suffer from side-channel attacks which could breach data privacy.
\begin{itemize}
    \item {\itshape Timing side channel \cite{gupta2016using}.} Given different inputs, the running time of the program in the enclave may be different too. By monitoring the difference over the execution time, an attacker may derive some sensitive information about the input, e.g., a secret value reflecting how many times a loop is executed. To solve this problem, one approach is to ensure the program loaded into the enclave always takes approximately the same amount of execution time. 
    \item {\itshape Memory side channel \cite{sasy2017zerotrace}.} In some programs, the access pattern to the non-PRM memory may depend on the enclave's secret. For example, if the enclave runs a binary search program, the access pattern is a path from the root node to a leaf node, which may reveal the real secret. Therefore the loaded program should remove any correlation between the memory access pattern and the insider secret. 
    \item {\itshape Network side channel \cite{zheng2017opaque}.} In a distributed system, network access patterns (e.g., sorting or hash partitioning) may reveal sensitive information. Therefore, the network access protocol should also be well-designed such that no sensitive information can be deduced from the patterns.
\end{itemize}

Second, TEEs may also be vulnerable to rewind attacks (or replay attacks), which allows an attacker to actively reset the state of the computation \cite{bellare2001identification}. Note that such attacks may result in catastrophic consequences for stateful computation, such as limited-attempt password checking \cite{skorobogatov2016bumpy}. We cannot always expect the TEE to maintain the state securely for two reasons. (1) TEE may be stateless in practice. 
(2) Even for stateful TEE, the limited protected memory makes it fail to correctly maintain a large state. Therefore, the designer should be careful and provides special mechanisms to counter rewind attacks.


\subsection{Our Contributions} \label{subs:contributions}
TEE provides a promising hardware-based alternative for achieving secure computation with a CPU-level performance, and thus bridges the gap between academic research and industrial adoption of secure computation. Trusted hardware provides a powerful abstraction for building secure systems and can be applied to many potential applications, such as machine learning algorithms \cite{ohrimenko2016oblivious,florian2019slalom,kunkel2019tensorscone,mo2020darknetz,ng2021goten,mo2021ppfl}, cloud computing \cite{santos2012policy, martignoni2012cloud, baumann2015shielding, schuster2015vc3, tamrakar2017circle, correia2020omega}, blockchain \cite{zhang2016town, lind2018teechain, bentov2019tesseract, bowe2020zexe}, network traffic analysis \cite{duan2019lightbox},  scientific computation \cite{shaon2017sgx}, data analytics \cite{zheng2017opaque, priebe2018enclavedb, kim2019shieldstore, zhou2021veridb, dave2020oblivious}, and privacy-preserving COVID-19 contact tracing \cite{kato2021pct}.

To help readers systemically perceive the core design of TEE-based secure computation protocols, this article presents a comprehensive survey on the state of the art of this research field. Our contributions are summarized as follows:
\begin{itemize}
    \item First, we propose two taxonomies for TEE-based secure computation protocols from the computing framework and TEE utilization perspectives. According to the framework, we categorize secure computation protocols into secure outsourced computation (SOC), secure distributed computation (SDC), and secure multiparty computation (SMC). 
    Based on the TEE utilization, the existing protocols are classified into three categories: (1) pure TEE based protocols, which use TEE as a black box, (2) TEE with oblivious primitives, which can prevent side-channel and leakage attacks, and (3) TEE with trusted ledge, which can resist attacks targeting TEE state.
    \item Second, comprehensive assessment criteria are presented for the purpose of evaluating and comparing the existing protocols. The assessment criteria are made from four dimensions: (1) \textit{setting} that specifies the computing function and framework; (2) \textit{methodology} that describes the TEE utilization method and the additional crypto-primitives used; (3) \textit{security} that highlights the security features of a protocol, such as privacy, integrity, fairness, and freshness; and (4) \textit{performance} that shows the experimental configurations and results.
    \item Third, comprehensive comparisons among the existing protocols are made in accordance with the above assessment criteria. Specifically, this article first compares the existing proposals for general-purpose secure computation. Then detailed comparisons are made on two specific computational tasks: (1) machine learning, which consists of training and inference; and (2) encrypted database queries.
    \item Lastly, we discuss several future research directions and challenges in the light of our observations on the state-of-the-art TEE-enabled secure computation solutions.
\end{itemize}

This survey also aims to help non-specialists comprehend TEE-enabled secure computation and assist practitioners to determine appropriate TEE technologies for system deployment. 

\section{Taxonomy and Assessment Criteria} \label{sec:criteria}
This section presents taxonomy and assessment criteria for TEE-based secure computation in Sec. \ref{subs:taxonomy_criteria} and in Sec. \ref{subs:assessment_criteria}, respectively.

\subsection{Taxonomy of TEE-based Secure Computation} \label{subs:taxonomy_criteria}

\begin{figure}[htpb]
  \centering
  \subcaptionbox{Outsourced Framework\label{subfig:outsourced}}{%
    \includegraphics[width=0.67\textwidth]{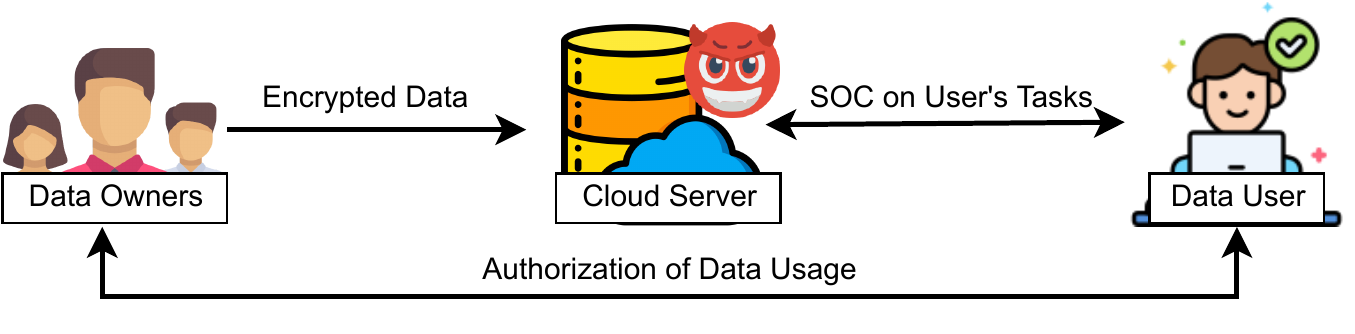}
  }
  \vspace{10pt}
  \subcaptionbox{Distributed Framework\label{subfig:distributed}}{%
    \includegraphics[width=0.67\textwidth]{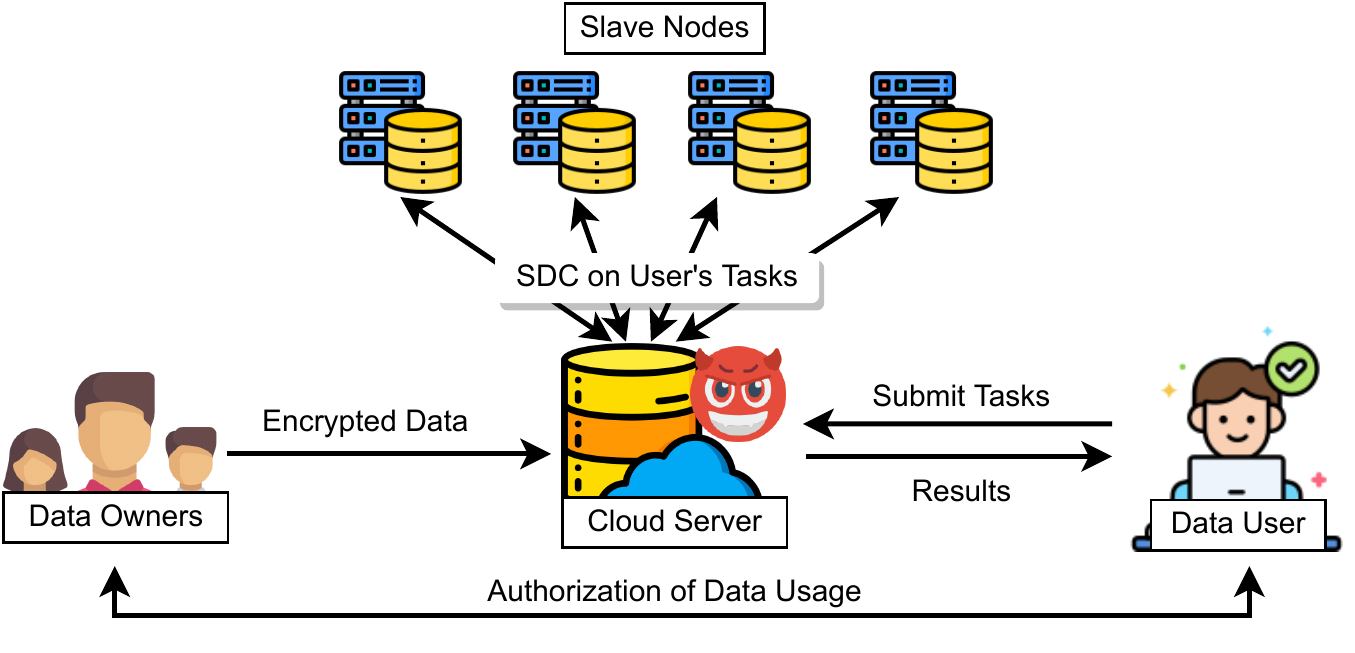}
  }
  \vspace{10pt}
  \subcaptionbox{Multi-Party Framework\label{subfig:multiparty}}{%
    \includegraphics[width=0.67\textwidth]{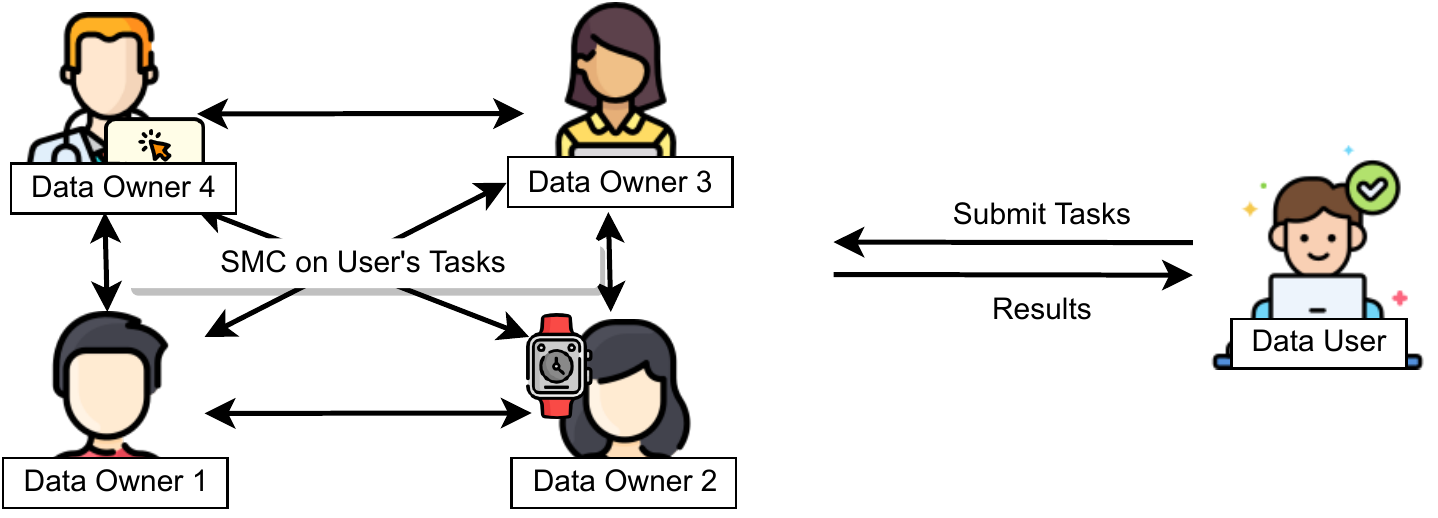}
  }
  \caption{Three Different Frameworks For Secure Computation}\label{fig:secureComputation}
  \Description{Three Different Frameworks For Secure Computation.}
\end{figure}

We categorize TEE-based secure computation protocols based on two orthogonal aspects: the computation framework and the TEE utilization.

Based on the computation framework, we separate TEE-based secure computation protocols into the following three categories:

\begin{itemize}
  \item \textbf{Secure outsourced computation (SOC).} Fig. \ref{subfig:outsourced} shows the framework of SOC, in which data owners outsource encrypted data to a single cloud server. Subsequently, an authorized data user will run an SOC protocol with the cloud server to realize desired computation tasks.
  In practice, the single cloud is an untrusted entity, which may behave as a passive/semi-honest attacker or an active/malicious attacker. 
  \item \textbf{Secure distributed computation (SDC).} Fig. \ref{subfig:distributed} shows the framework of SDC, in which the encrypted data is outsourced to multiple servers. Specifically, one master node is in charge of coordinating the computational tasks, and several slave nodes execute the computation in a distributed manner. An SDC protocol is run between an authorized data user and these server nodes. Specifically, each slave node calculates an intermediate result, and the master node responds data user with the final output, which is reconstructed from intermediate results. The master node and slave nodes may be semi-honest or malicious. Compared to SOC, the interactions among the servers and the intermediate results in SDC expose a larger attack surface to attackers.
  \item \textbf{Secure multi-party computation (SMC).} Fig. \ref{subfig:multiparty} shows the framework of SMC, in which data owners jointly run a SMC protocol themselves without losing control of their data. The difference between SMC and SDC is that SMC does not involve a master node to coordinate computation tasks. 
  The data owners (also called parties) participating in the computation don't trust each other. Specifically, each party in SMC tries to protect their sensitive information from other parties who may behave semi-honestly or even maliciously (i.e., by undermining the protocol).
\end{itemize}

Based on TEE utilization, i.e., whether it is used alone or in conjunction with other tools for preventing side-channel or state-related attacks, we can also categorize the protocols  into the following types:

\begin{itemize}
	\item \textbf{TEE as a black box.} TEE enables general-purpose computation with confidential and integrity guarantee against eavesdropping adversaries. Because the input and output are encrypted by an authenticated encryption scheme (e.g., AES-GCM), the untrusted host cannot learn sensitive information. Many existing solutions regard TEE as a black box and fall in this category. If a protocol requires a stateful hardware device, we mark it as ``stateful TEE''; otherwise, we mark is as "stateless TEE"..
	\item \textbf{TEE with oblivious primitives.} As mentioned earlier, although a host cannot directly access TEE's state, sensitive information may still be leaked via different side-channels. To prevent such leakage, well-designed oblivious primitives (OP) \cite{sasy2019oblivious} and differential obliviousness \cite{mazloom2018secure,xu2018using,allen2019algorithmic} are leveraged to mitigate side-channel attacks from different levels. The resulting proposals, such as ZeroTrace \cite{sasy2017zerotrace} and ObliJoin\cite{chang2022towards}, not only achieve confidentiality and integrity but also avoid sensitive information leakage from other channels. The protocols in this category are labeled as ``stateful TEE + OP'' or ``stateless TEE + OP'' for comparison.
	\item \textbf{TEE with trusted public ledger.} When TEE is deployed on a remote malicious host, the latter may submit illegal inputs to the enclave. For example rewind attack in a stateful computation, the host may provide a stale encrypted state to the enclave, which may lead to catastrophic effects. Therefore a trusted public ledger (PL) \cite{kaptchuk2017giving} can be  introduced to ensure the legitimacy of inputted state, 
	and the protocols that fall in this category are labeled as "stateful TEE + PL" or "stateless TEE + PL" for comparison.
\end{itemize}

\subsection{Assessment Criteria} \label{subs:assessment_criteria}

\begin{figure}[ht]
  \centering
  \includegraphics[width=0.4\linewidth]{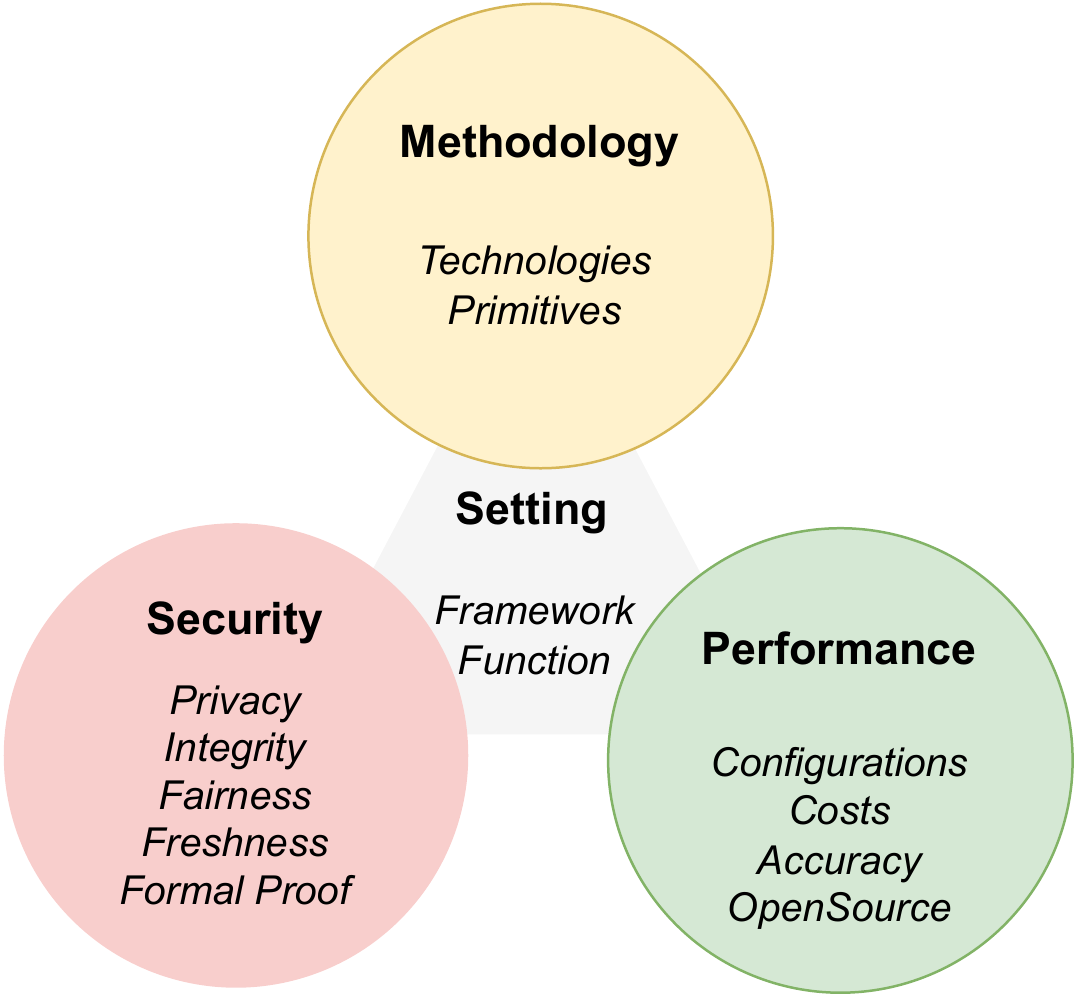}
  \caption{Assessment Criteria}\label{fig:criteria}
  \Description{Assessment}
\end{figure}

For a systematic and fair comparison of existing proposals for TEE-based secure computation, we present the common assessment criteria with respect to four aspects: setting, methodology, security and performance (shown in Fig. \ref{fig:criteria}). These criteria are applied in the comparison of existing protocols throughout this survey.

\subsubsection{Setting} \label{subsubs:setting}
Secure computation protocols are designed for different purposes and can follow different computation frameworks. For achieving a fair comparison, we distinguish the secure computation setting from two aspects: {\itshape framework} and {\itshape function} as shown in Fig. \ref{fig:criteria}. 
The framework specifies the architecture of the designed protocol, which has been elaborated in Sec. \ref{subs:taxonomy_criteria}. The function represents the primary goal of the protocol, such as a machine learning task (training or inference), scientific computation (such as matrix multiplication and polynomial evaluation), data analytics (such as SQL queries and statistical analysis), or a general-purpose function.

\subsubsection{Methodology} \label{subsubs:methodology}
Methodology describes the design principle of a proposal and will guide developers to facilitate the practical deployment. 
We describe the methodology of TEE-based protocols from two aspects (shown in Fig. \ref{fig:criteria}): TEE {\itshape utilization} and crypto {\itshape primitives}. The former describes how TEE is used in a protocol (Sec. \ref{subs:taxonomy_criteria}) and the latter presents which cryptographic toolkit is leveraged.
It is worth noting that many crypto primitives are employed to deal with the security issues outside the TEE, such as authenticated encryption, digital signatures, secret sharing, and zero-knowledge proof. Therefore, we also highlight crypto primitives in the comparison.

\subsubsection{Security} \label{subsubs:security}
Generally, security evaluation consists of two dimensions: {\itshape security goals} and {\itshape adversarial models}. According to system requirements, different security goals are defined under different adversarial models. We present the common security goals and adversarial models as follows. 
\begin{itemize}
    \item \textbf{Security goals.} Security goals define which security requirements should be achieved in a secure computation protocol. In this article, four types of security goals are considered: privacy, integrity, fairness, and freshness. Specifically,
        \begin{itemize}
            \item {\itshape Privacy.} Privacy in secure computation means that sensitive data (including input, intermediate result, output, or other metadata during computation such as access pattern) is not visible to potential attackers. Moreover, the sensitive data may include the concrete queries in privacy-preserving database queries or model parameters and inference data in privacy-preserving machine learning. All sensitive information should be safeguarded from being leaked to any unauthorized party (e.g., a cloud server). 
            \item {\itshape Integrity.} Integrity in secure computation means that no malicious party can sabotage the protocol execution and produce the wrong output for other parties. 
            For example, in federated machine learning, a party may launch a backdoor attack to corrupt the performance of the training model by training its local model maliciously \cite{yin2021comprehensive}.
            \item {\itshape Freshness.} Freshness in secure computation means that the involved party cannot leverage stale state or old data for stateful computation. For example, an adversary may launch a  replay attack by re-transmitting stale valid messages to honest parties, thereby fooling the honest parties. Therefore, any messages transmitted between the parties should be passed with freshness checking, which ensures the state or the data is up to date.
            \item {\itshape Fairness.} Fairness is a natural and desirable security requirement in SMC, which has shown its importance in auctions, contract signing, payment and other application scenarios. Fairness ensures that either all parties receive the protocol output or no one does \cite{choudhuri2017fairness}. While complete fairness for general computation is impossible \cite{cleve1986limits}, several possible and useful fairness solutions have been proposed such as gradual release \cite{blum1983exchange}, probabilistic fairness \cite{luby1983simultaneously}, optimistic exchange \cite{bao1998efficient}, fairness with penalties \cite{andrychowicz2014secure}, and $\Delta$-fairness \cite{pass2017formal}.
        \end{itemize}
    \item \textbf{Adversarial models.} Adversarial models abstract attackers' power. In general, secure computation protocols consider two types of attackers: semi-honest and malicious. Specifically,
        \begin{itemize}
            \item {\itshape Semi-honest.} A semi-honest adversary, also called honest-but-curious adversary, is assumed to follow the protocol's instructions honestly so that the tasks are conducted and completed correctly. However, the attacker is curious and tries to learn sensitive information from each step of the computation.
            \item {\itshape Malicious.} A malicious adversary is stronger than the semi-honest one since it can arbitrarily deviate from the prescribed protocol. A malicious adversary could learn more information about the sensitive data through active attacks or save computation or storage costs by performing lazy computation and returning random results to other honest parties.
        \end{itemize}
\end{itemize}

Besides, formal proof is an important aspect of evaluating security. A formal abstraction of the secure enclave was introduced in \cite{pass2017formal}, which is also adopted by the subsequent works with formal security analysis, such as \cite{choudhuri2017fairness, paul2019efficient, wu2022hybrid}. We present the formal TEE abstraction (also denoted $\mathcal{G}_{att}$) in our supplementary material (Sec. B).

\subsubsection{Performance} \label{subsubs:performance}
We present the performance assessment criteria from four aspects: experimental configuration, costs, accuracy, and open-sourceness, which are elaborated below. We should note that a fair comparison of secure computation protocols is not a trivial task, because the performance indicators may be obtained from different data sets and be exhibited in many different manners. Thus, we are not able to compare all the protocols under the same experimental configuration. 
Instead, we aim to provide a relatively objective comparison based on the following four aspects.
\begin{itemize}[leftmargin=10pt]
    \item \textbf{Experimental configuration.} Experimental configurations are diverse, such as computer specifications, platforms, workloads, data sets, and network architectures. Therefore, we will specify the necessary experimental configurations in the comparison, which would also help readers to understand the performances in real deployment.
    \item \textbf{Costs.} The secure computations exhibit their performances in many different ways, such as complexity estimation without experiments, absolute time consumption, a percent indicating relative cost, throughput, or others. To be as unified as possible, we present its baseline in the comparison. Specifically, the following baselines are considered.
        \begin{itemize}
            \item {\itshape Plaintext model.} The plaintext model means that the function is computed outside the TEE without any protection, which exhibits exceptionally high performance. For comparing with the plaintext model, the protocols' performance will be worse, and we highlight the ``slowdown'' factor in comparisons.
            \item {\itshape Fully in enclave.} The fully in enclave model means that the entire computational task is completed in the enclave, which may trigger the page swapping of the TEE and thus results in a large TCB size and worse performance. A well-designed protocol's performance will usually be better than the fully in enclave model, and we highlight the ``speedup'' factor in comparisons.
            \item {\itshape State of the arts.} Some references compare their protocol with the state of the arts. In comparison, we highlight ``slowdown'' or ``speedup'' factors for degrading or improvement in performance.
        \end{itemize}
    \item \textbf{Accuracy.} Accuracy is an essential criterion in many machine learning algorithms. Informally, accuracy is the fraction of predictions the model got right, and high accuracy is preferred. Accuracy in general secure computation means that the absolutely correct results should be returned. However, differential privacy based solutions adds noise perturbations to the results for meeting specific privacy requirements and thus may not produce absolutely correct results.
    \item \textbf{Open-Sourceness}. Last but not least, we also summarize whether an implemented protocol is open-sourced, which helps other researchers to understand the proposal better and to check its validity and performance.
\end{itemize}

\subsection{Roadmap}

\begin{figure}[ht]
  \centering
  \includegraphics[width=1.05\linewidth]{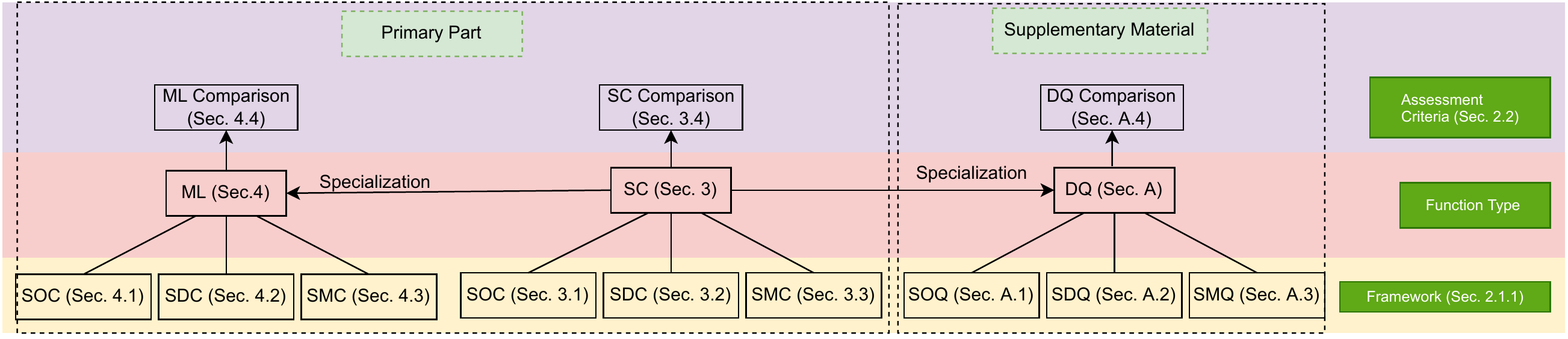}
  \caption{Taxonomy, Assessment, and Organizations.}\label{fig:taxonomy}
  \Description{Taxonomy and organizations.}
\end{figure}

Fig. \ref{fig:taxonomy} presents the roadmap of this survey. The completed survey consists of the primary part and additional supplementary material. In the primary part, TEE-based secure computation protocols for general-purpose function and machine learning are sequentially reviewed in Sec. \ref{sec:secure_computation} and Sec. \ref{sec:machine_learning}, respectively. In the supplementary material, we review the TEE-based secure computation protocols for database queries in Sec. A.
In each section, we present problem statements and research status with regard to different frameworks. For example, Sec. \ref{sec:secure_computation} reviews TEE-based protocols for general-purpose secure computation. Following the taxonomy presented in Sec. \ref{subs:taxonomy_criteria}, we present the SOC, SDC, and SMC in Sec. \ref{subs:soc}, Sec. \ref{subs:sdc}, and Sec. \ref{subs:smc}, respectively. After that, we provide thorough comparisons in Sec. \ref{subs:smc_comparisons} according to assessment criteria shown in Sec. \ref{subs:assessment_criteria}.
As two important and special research directions in secure computation, we organize contents for Machine Learning (in Sec. \ref{sec:machine_learning}) and Database Queries (in Sec. A) in a similar way as in Sec. \ref{sec:secure_computation}.
Finally, concluding remarks and future directions are made in Sec. \ref{sec:conclusion}.

\section{Secure Computation Using TEEs}\label{sec:secure_computation}

In this section, we turn to discuss and compare existing general-purpose secure computation protocols using TEEs. According to different frameworks, we successively present secure outsourced computation, secure distributed computation, and secure multiparty computation.

\subsection{Secure Outsourced Computation}\label{subs:soc}

\subsubsection{Problem Statement}

Secure outsourced computation (SOC) allows a resource-limited client to outsource a computation task $(\Phi, x)$ to a powerful cloud server, where $\Phi$ is a function or a program for computation, and $x$ is a sensitive input. To provide privacy guarantee, the task $(\Phi, x)$ should be encrypted to $(\enc{\Phi}, \enc{x})$ and the cloud server finishes the computation in encrypted domain, $(\enc{y}, \Gamma) = \text{SOC}(\enc{\Phi}, \enc{x})$, where $\enc{y}$ is encrypted output and $\Gamma$ is the proof for checking the integrity of execution. Upon receiving $\enc{y}$, the client decrypts it to the final output $y$. The {\itshape consistency} of SOC means that the final output $y = \Phi(x)$. The {\itshape privacy } means that the cloud server cannot learn any sensitive information about $\Phi$ (function privacy), $x$ (input privacy), or $y$ (output privacy). The {\itshape integrity} means that the server correctly executes the task $(\Phi, x)$.

\begin{figure}[h]
  \centering
  \includegraphics[width=0.6\linewidth]{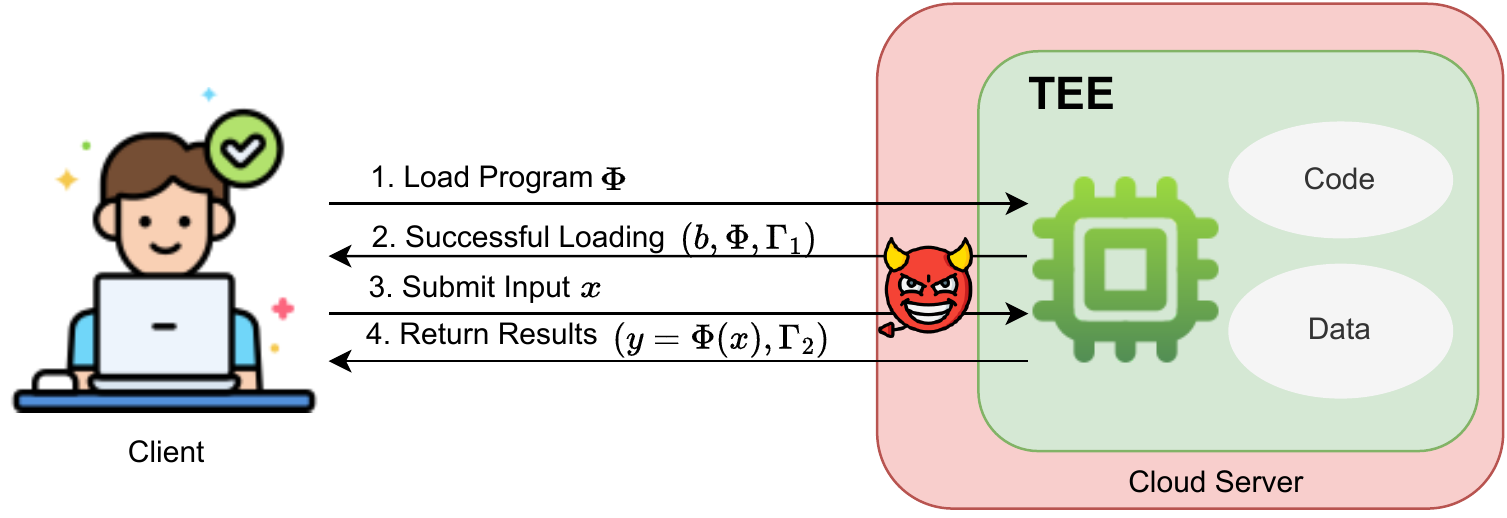}
  \caption{Secure Outsourced Computation}\label{fig:SOC}
  \Description{Assessment}
\end{figure}

Shan et al. \cite{shan2018practical} surveyed the non-TEE approaches for SOC, such as fully homomorphic encryption, additive homomorphic encryption, random transformation, secret sharing, etc. 
Since the arising of trusted hardware, the TEE-based methods received a lot of attentions due to its high performance.
As shown in Fig. \ref{fig:SOC}, we provide the basic SOC workflow to explain the interactions between a client and a TEE-equipped server. The client builds a secure channel between the enclave by remote attestation, after which four steps are followed. (1) The client loads the program $\Phi$ into the enclave, and (2) The enclave responds with a triple $(b, \Phi, \Gamma_1)$, in which $\Gamma_1$ proves the program is loaded successfully ($b=1$); otherwise, $b=0$ and $\Gamma_1 = \bot$. (3) The client submits an input $x$ to the enclave. (4) The enclave returns the final results $(y, \Gamma_2)$, in which $y = \Phi(x)$ and $\Gamma_2$ proves the program is executed honestly. Note that all messages between client and enclave are sent to each other over the secure channel.

\subsubsection{Research Status}

In \cite{barbosa2016foundations}, Barbosa et al. presented the notion {\itshape attested computation}, which focuses on integrity assurance by building a secure channel via key exchange. With their proposed {\itshape utility theorem}, they presented the first SOC framework using trusted hardware, which also offers formal security analysis by simulation-based proof. Another formal treatment for TEE-based SOC is the symbolic model \cite{jacomme2017symbolic, subramanyan2017formal}, which aims to design secure remote attestation execution protocols. To against side-channel attacks, ObliCheck \cite{son2021oblicheck} employs symbolic execution to check whether all execution paths exhibit the same observable behavior. In \cite{pass2017formal}, Pass et al. presented the {\itshape formal abstractions and rigorous exploration} of TEE and offered a TEE-based stateful obfuscation to show the power of attested execution. Then, for the first time, they described the basic $\mathcal{G}_{att}$ abstraction (Sec. B in supplementary material) capturing the essence of TEE. After that, Pass et al. employed the abstraction to design an SOC protocol, which was proved to be UC-secure in $\mathcal{G}_{att}$-hybrid model. Following \cite{pass2017formal}, many works in hardware-supported secure computation protocols were formally treated and proved to be UC-secure by employing $\mathcal{G}_{att}$ abstraction (e.g., \cite{wu2022hybrid}).

To mitigate rewind attack, a public ledger (PL) served as a trusted data source is utilized to ensure the TEE state freshness while computing a stateful function \cite{kaptchuk2017giving}. The authors proposed a hybrid computing framework in which each input is registered from the public ledger. The public ledger signs the inputs by a signature scheme, and then the enclave checks the freshness of the state according to the public ledger before executing the general function computation. Therefore, it achieves stateful, interactive functionality, even on devices without persistent storage (such as mobile devices).

The fairness in SOC is different from the one in SMC. In \cite{dang2019towards}, Dang et al. explored the fair exchange in a marketplace for secure outsourced computation. In the market, the client first commits a remuneration and then submits a computational task to computing nodes. The computing nodes get the remuneration if the task is completed successfully. To achieve the goals, the authors proposed a framework Kosto, which introduced a third party (called a broker) to assist the clients in attesting correct instantiation of the enclaves and computing nodes' certain commission fees.

\subsection{Secure Distributed Computation} \label{subs:sdc}

\subsubsection{Problem Statement}

Secure distributed computation (SOC) is run between a master node and multiple slave nodes. Generally, each slave node computes an intermediate result over its inputs, and then the master node aggregates all intermediates into the final results. Specifically, the task $(\Phi, x)$ is split into many sub-computational tasks $(\Psi, x_i)$, $i = \{1, 2, \cdots, N\}$ and an aggregation task $\Pi$ such that
\begin{equation}
\Phi(x) = \Pi\left(\Psi(x_1),\Psi(x_2), \cdots, \Psi(x_N)\right) 
\end{equation}
Each slave node runs a sub-computational task, and the master node takes charge of the aggregation task. According to different secure computation protocols, the algorithms $\Psi$ and $\Pi$ may work interactively. Note that privacy and integrity also should be achieved in an SDC protocol.

\begin{figure}[h]
  \centering
  \includegraphics[width=\linewidth]{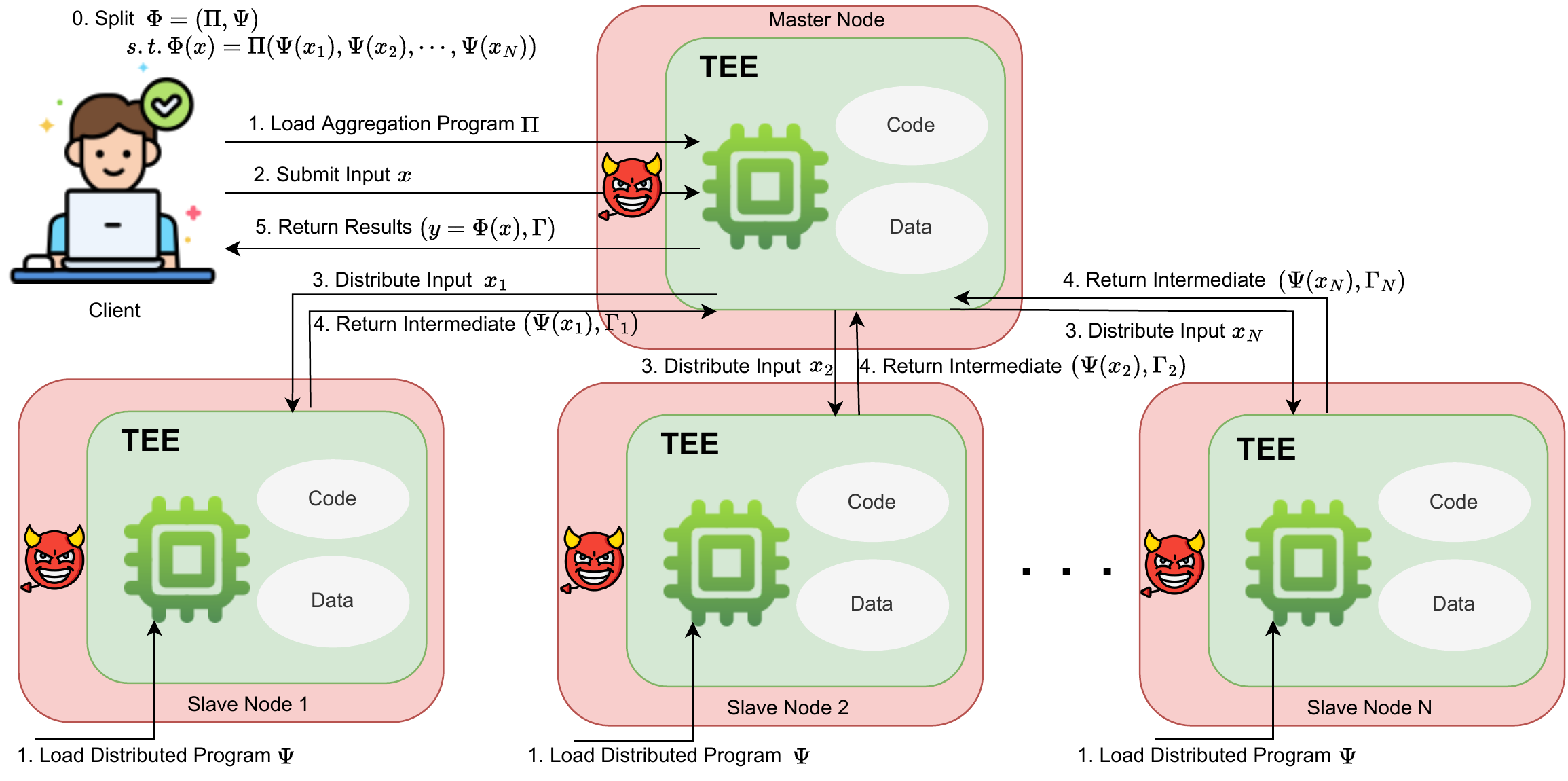}
  \caption{Secure Distributed Computation}\label{fig:SDC}
  \Description{Assessment}
\end{figure}

Shyuan et al. \cite{ng2020survey} surveyed the non-TEE approaches for SDC, such as linear secret sharing, homomorphic encryption, and random transformation, etc. These solutions may be used on different platforms such as MapReduce and Spark. Unlike the above approaches, we survey the hardware-supported technologies for SDC protocols. As shown in Fig. \ref{fig:SDC}, we provide the basic SDC workflow to explain the interactions between the master node and multiple slave nodes. 
(1) The client builds secure channel with servers (including master and slave nodes) via remote attestation, and then ensures distributed program $\Psi$ and aggregation program $\Pi$ are successfully loaded into the slave nodes' enclave and the master node's enclave, respectively.
(2) The client submits an input $x$ to the master node. (3) The master node splits the inputs $x$ into $x_1, x_2, \cdots, x_N$ and then distributes $x_i$ to $i$-th slaves. (4) Each slave node finishes the sub-computational tasks in TEE and responds to the master node with a tuple $\left(\Psi(x_i), \Gamma_i\right)$, in which $\Psi(x_i)$is the $i$-th intermediate result and $\Gamma_i$ proves slave node executes the program $\Psi$ honestly. (5) Finally, the master node aggregates all intermediate results by $\Pi$ and then sends the final results $(y, \Gamma)$ to the client, in which $\Gamma$ proves the program $\Pi$ is honestly executed. 
Note that all messages among the entities are exchanged over the secure channel.

\subsubsection{Research Status}

In \cite{schuster2015vc3}, a TEE-based framework VC3 for secure distributed computation in the cloud was proposed by Schuster et al. VC3 enables users to finish computations in the distributed MapReduce setting while maintaining the confidentiality of data and ensuring the accuracy and completeness of the final results. VC3 does not modify the original Hadoop and thus maintains compatibility with Hadoop. In the design of VC3, the trusted computing base (TCB) excludes the operating system and the hypervisor; therefore, it achieves confidentiality and integrity even if the host is corrupted. Relying on the well-designed safe r/w operators inside the enclave to unsafe memory, VC3 facilitates the deployment of new SDC protocols under the MapReduce platform. However, it does not solve the access pattern leakage and does not support multi-user key exchange, which may result in huge overhead in the remote attestation. On top of VC3, Ohrimenko et al. in \cite{ohrimenko2015observing} observed that intermediate traffic might leak sensitive information by employing geographical data as the motivating example. They proposed a mechanism to hide the correlation between the map function and the reduce function against a Map-Reduce game and then prove security by a heuristic method.

To overcome the limited memory of TEEs, an ORAM controller is installed insider the TEE and an ORAM storage is arranged in external memory or disk. By contrast, M2R proposed in \cite{dinh2015m2r} focuses on resisting the side-channel attacks in SDC where each unit is equipped with trusted hardware. M2R allows adversaries to obtain channel information and launch passive attacks (dataflow patterns, order of execution, time of access) and active attacks (tuple tampering, misrouting tuples). While expensive ORAM introduces a multiplication factor $O(\log{N})$ overhead to the latency, M2R incurs only an additive logarithmic factor by grouping then shuffling technique, which employs the shuffler in the MapReduce platform directly without degrading privacy. However, all the above schemes only work on small-scale data (e.g., 1-5GB).

To build a UC-secure SDC protocol, ObliDC \cite{wu2019oblidc} is an SGX-based framework that introduces the ODC-privacy to capture the privacy leakage in the entire life circle, which is modeled as a data-flow graph. Specifically, it decomposes the distributed computing into four phases: job deployment, job initialization, job execution, and results return, and for each step, a UC-secure two-party protocol was designed to meet ODC privacy. It is worth mentioning that obliDC achieves both semantic security and oblivious traffic even if the adversaries are malicious. Furthermore, ObliDC supports any general-purpose computation by defining the sensitive C-style code to be run in an enclave.

\subsection{Secure Multiparty Computation} \label{subs:smc}

\subsubsection{Problem Statement}

Secure multiparty computation (SMC) is aimed at computing a joint task among a number of distinct yet connected parties while revealing nothing but the output. Specifically, assume each participant has an input $x$. In SMC protocol for a task $\Phi$, the correctness means that $\Phi(x_1, x_2, \cdots, x_n)$ is computed correctly, and the privacy means that the execution of the protocol only reveals the final result to each party. Integrity ensures that all parties follow the protocol steps honestly. Some applications may enforce the SMC to be fair, which means that malicious parties receive the outputs if and only if other honest parties also receive their outputs. SMC has shown its importance in many application domains, such as privacy-preserving biding, private DNA comparison, privacy-preserving machine learning, etc. 

Choi et al. \cite{choi2019secure} surveyed the non-TEE approaches for SMC.
Since Yao's Garbled Circuits \cite{yao1986generate}, many researchers have targeted designing a secure and efficient SMC protocol. Nevertheless, the solution is still impractical because of the huge communication or extensive computational cost. Many existing non-hardware solutions focus on improving efficiency, such as Sharemind \cite{bogdanov2008sharemind}, VMCrypt \cite{malka2011vmcrypt}, ABY \cite{demmler2015aby}, SPDZ \cite{damgaard2012multiparty}, HyCC \cite{buscher2018hycc}, etc. Most of them rely on the primitives such as GC-OT, GMW, FHE, Secret Sharing, ORAM, etc. In this survey, we review the approaches relying on TEE.

\begin{figure}[h]
  \centering
  \includegraphics[width=0.7\linewidth]{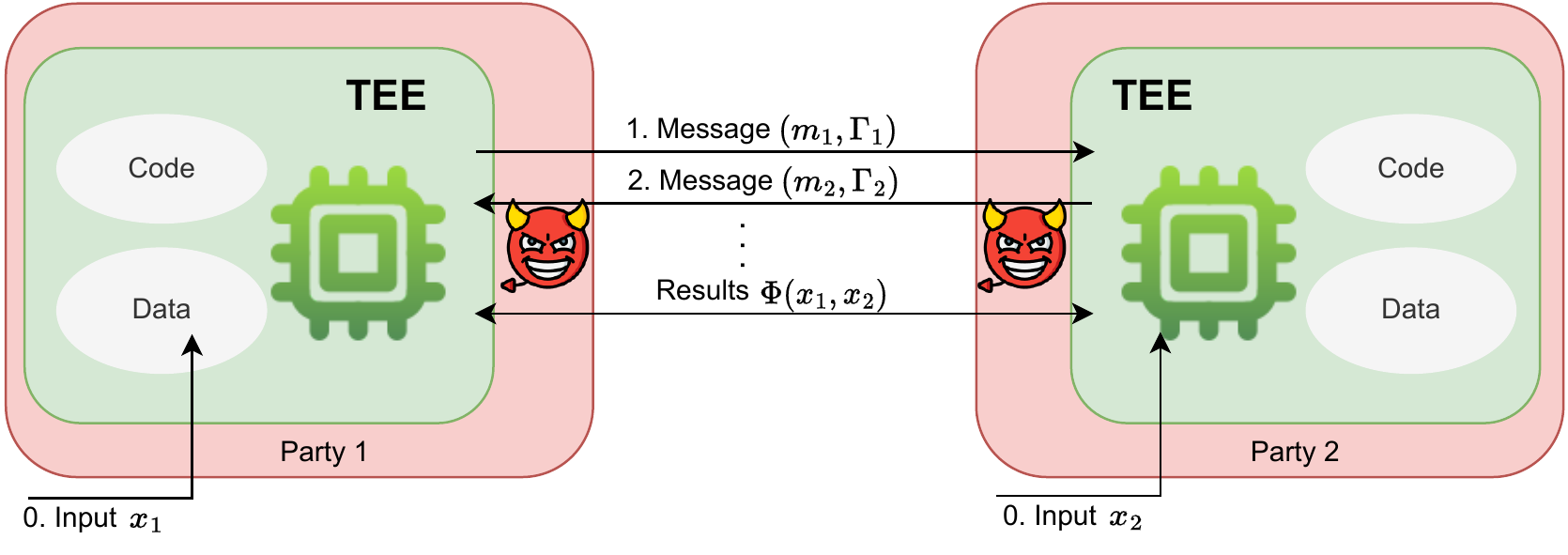}
  \caption{Secure Two Party Computation}\label{fig:SMC}
  \Description{Assessment}
\end{figure}

Secure two-party computation workflow is presented in Fig. \ref{fig:SMC}, which can be easily extended to multiple parties. Both parties equip TEEs and build a secure channel via remote attestation. Taking their own data $x_1$ and $x_2$ as inputs, two parties finish the computation in an interactive manner, which consists of multiple rounds of communication.
At each round, a message associated with a proof $(m_i, \Gamma_i)$ is sent to another party where $m_i$ is a committed message to ensure the privacy of the execution, and $\Gamma_i$ is employed to prove that $m_i$ is computed honestly. Finally, the two parties may run a fair protocol to distribute the final result $\Phi(x_1, x_2)$. Note that all messages between the two parties are sent over the secure channel.

\subsubsection{Research Status}

In \cite{gupta2016using}, Gupta et al. showed the naive execution of functions within TEE might result in significant side-channel leakages, such as runtime and RAM access patterns. Instead of relying on SGX enclaves to process the entire computation, the authors leveraged the garbled circuits to handle the sensitive portion of the computation and advocate separating the computation into separate circuits and SGX components. In such a way, the secrets are protected even if the enclave is compromised. However, they did not treat the SMC problem formally. Bahmani et al. in \cite{bahmani2017secure} extended the attested computation \cite{barbosa2016foundations} to labeled attested computation (LAC), which makes each party a label. Thus users can get attestations of parts of code corresponding to specific labels. With the proposed LAC, they built an efficient SMC protocol from SGX by composing attested key exchange procedures for each participant in parallel. Furthermore, the TEE-based SMC protocol was proved by simulation, and some benchmarks compared to ABY \cite{demmler2015aby} are presented to show its efficiency.

Based on the $\mathcal{G}_{att}$-hybrid abstraction \cite{pass2017formal}, Pass et al. drew several surprising results on SMC. (1) UC-secure SMC protocol is impossible if at least one party is not equipped with a TEE. However, global Augmented Common Reference String (ACRS) \cite{canetti2007universally} model makes the UC-secure SMC realizable even if only one party is equipped with a TEE because it allows simulating corrupted clients during proof. (2) Combined with secure key exchange, UC-secure 2-party computation is possible when both parties have TEEs. (3) Combined with secure key exchange, if both parties have TEEs and trusted clocks, then fair 2-party computation is achievable. (4) Fairness is impossible for general functionalities when one party is not clock-aware, but fairness for certain functions would be possible with the help of the ACRS model.

To alleviate the burden of trust on the enclave, Choi et al. \cite{choi2019hybrid} explored a balanced method to split a function $f$. Instead of evaluating the entire function $f$ in the enclave, the proposed 2P-SFE protocol allows the designer to choose which components should be evaluated within the enclave. 2P-SFE modeled the enclave as a black box and was applied to two practical problems: private queries to a database and private navigation based on Dijkstra's algorithm. However, the scalability of 2P-SFE is limited. 
While 2P-SFE only works for boolean circuits, the authors in \cite{felsen2019secure} proposed a method to evaluate the universal circuit via Intel SGX, where the universal circuit can emulate any function by programming input bits to a given size. However, both \cite{choi2019hybrid} and \cite{felsen2019secure} assume the parties are semi-honest.
In \cite{wu2022hybrid}, Wu et al. proposed a generic framework HYBRTC, which introduces the concept of TEE trust levels. Specifically, in an SMC protocol, a party might fully trust TEE, while another party may only trust the TEE partially or never trust it. HYBRTC solves the problem when different levels of trust exist in an SMC protocol. From a high-level overview, for parties who think TEEs are reliable, their private inputs can be processed in the enclave. Otherwise, their private inputs should be protected by a cryptographic SMC protocol. Besides, HYBRTC also eliminates side-channel attacks.

\noindent \textbf{Fairness in SMC.} TEE was also explored to design fair SMC protocols. In \cite{choudhuri2017fairness}, Choudhuri et al. presented a hybrid model for achieving {\itshape fairness} by using a public ledger, even in the case of a dishonest majority. Note that existing solutions such as blockchains or Google’s certificate transparency logs can implement the public ledger. What important is that the proposed SMC has complete fairness, which is different from the method of penalizing an aborting party or achieving weaker notions such as $\Delta$-fairness. After that, \cite{paul2019efficient} presented a robust construction that ensures fairness, although it allows to output an incorrect value. While implementing fairness by leveraging a public ledger, the input, output and the state need to be stored on the chain, which brings the burden of the blockchain. In \cite{sinha2019luciditee}, LucidiTEE explored a method that does not store inputs, outputs or states on the chain. Besides, LucidiTEE also suggested integrating the history-based policy into SMC, which achieves fair delivery and enforces policy-restricted input. 

\begin{landscape}
\begin{table}[t]
  \caption{Comparisons of Secure Computation}
  \begin{threeparttable}
    \setlength\tabcolsep{1.2pt}
\renewcommand{\arraystretch}{1.8}
\tiny
\begin{tabular*}{\linewidth}{ccccccccccccc}
	\toprule
	\multirow{3}*{\scriptsize Schemes} & \multicolumn{2}{c}{\multirow{2}*{\scriptsize Settings}} & \multicolumn{2}{c}{\multirow{2}*{\scriptsize Methodology}} & \multicolumn{6}{c}{\scriptsize Security} &  \multicolumn{2}{c}{\multirow{2}*{\scriptsize Performance}} \\ \cmidrule(lr){6-11} 
				&   &   &   &   & \multicolumn{4}{c}{Security Goals} & \multirow{2}*{\makecell[c]{Adversarial\\Models}} & \multirow{2}*{\makecell[c]{Formal\\proof}} &   & \\ \cmidrule(lr){2-3} \cmidrule(lr){4-5} \cmidrule(lr){6-9} \cmidrule(lr){12-13}  
				& Framework & Function & Technologies & Primitives & Privacy & Integrity & Fairness & Freshness & & & Configurations & Costs \\ \midrule
	{[M2R \cite{dinh2015m2r}, USENIX Security, 2015]}            & SDC, n & \makecell[c]{Key-value Pair\\Analysis} & Stateless TEE + OP & \makecell[c]{PRP, Sig,\\PKE, SecMix} & IO, pattern, IR & \makecell[c]{IO, exe,\\Completeness} & \ding{55} & \ding{55} & \makecell[c]{Malicious Servers\\ HBC client} & \ding{55} & {\fontsize{4.5}{5.5} \selectfont \makecell[c]{Baseline: Fully in Enclave\\ Data size : 2-4GB}} & {\fontsize{4.5}{5.5} \selectfont 0.17-1.0x slowdown} \\ \hdashline[0.1pt/0.6pt]
	{[VC3 \cite{schuster2015vc3}, SP, 2015]}                     & SDC, n & General & Stateless TEE & \makecell[c]{PRF, Sig,\\PKE, KE} & IO, IR & \makecell[c]{IO, exe,\\Completeness} & \ding{55} & \ding{51} & \makecell[c]{Malicious Servers\\ HBC client} & \ding{55} & {\fontsize{4.5}{5.5} \selectfont \makecell[c]{Baseline: Plaintext Model\\ Datasize: 1GB}} & {\fontsize{4.5}{5.5} \selectfont WordCount: 0.05-0.4x slowdown} \\ \hdashline[0.1pt/0.6pt]
	{[OCF \cite{ohrimenko2015observing}, CCS, 2015]}             & SDC, n & \makecell[c]{Key-value Pair\\Analysis} & Stateless TEE + OP & \makecell[c]{PRF, Sig, PKE,\\KE, PRP} & IO, pattern, IR & \makecell[c]{IO, exe,\\Completeness} & \ding{55} & \ding{55} & \makecell[c]{Malicious Adaptive\\Servers\\ HBC client} & \ding{55} & {\fontsize{4.5}{5.5} \selectfont \makecell[c]{Baseline: Plaintext Model\\ Datasize: 2GB}} & {\fontsize{4.5}{5.5} \selectfont Aggregate: 2x slowdown} \\ \hdashline[0.1pt/0.6pt]
	{[2P-SFE \cite{gupta2016using}, FC, 2016]}                   & SDC, 2 & Stateless Function & Stateless TEE & GC & IO & exe & \ding{55} & \ding{55} & Malicious Server & \ding{55} & \multicolumn{2}{c}{\ding{55}} \\ \hdashline[0.1pt/0.6pt]
	{[Att. Comp. \cite{barbosa2016foundations}, EuroS\&P, 2016]} & SOC    & General & Stateful TEE & KE, AE & IO, state & IO, exe & \ding{55} & \ding{55} & Malicious Server & SIM & \multicolumn{2}{c}{\ding{55}} \\ \hdashline[0.1pt/0.6pt]
	{[Labeled Att.Comp. \cite{bahmani2017secure}, FC, 2017]}     & SMC, n & General & Stateful TEE & KE, AE & IO, state & IO, exe & \ding{55} & \ding{51} & Malicious Parties & SIM & {\fontsize{4.5}{5.5} \selectfont Baseline: ABY} & {\fontsize{4.5}{5.5} \selectfont \makecell[c]{HD: 5-10x speedup\\ PSI: 8-200x speedup}} \\ \hdashline[0.1pt/0.6pt]
	{[PST17-I \cite{pass2017formal}, EUROCRTPT, 2017]}           & SOC    & General & Stateful TEE & Sig, AE & IO & IO, exe & \ding{55} & \ding{51} & Malicious Server & UC & \multicolumn{2}{c}{\ding{55}} \\ \hdashline[0.1pt/0.6pt]
	{[PST17-II \cite{pass2017formal}, EUROCRTPT, 2017]}          & SMC,2  & General & Stateful TEE & \makecell[c]{Sig, PKE,\\AE, NIZK} & IO & IO, exe & \ding{55} & \ding{51} & Malicious Parties & UC & \multicolumn{2}{c}{\ding{55}} \\ \hdashline[0.1pt/0.6pt]
	{[PST17-III \cite{pass2017formal}, EUROCRTPT, 2017]}         & SMC,2  & General & Stateful TEE & \makecell[c]{Sig, PKE, AE,\\NIZ, Clock} & IO & IO, exe & $\Delta$-Fairness & \ding{51} & Malicious Parties & UC & \multicolumn{2}{c}{\ding{55}} \\ \hdashline[0.1pt/0.6pt]
	{[FSMC \cite{choudhuri2017fairness}, CCS, 2017]}             & SMC, n & General & Stateful TEE + PL & \makecell[c]{OWF, Sig,\\AE, Com} & IO & IO, exe & \ding{51} & \ding{55} & Malicious Parties & UC & \multicolumn{2}{c}{\ding{55}} \\ \hdashline[0.1pt/0.6pt]
	{[2P-SFE\tnote{$\dagger$} \cite{hunt2018chiron}, AsiaCCS, 2018]}              & SMC, 2 & Garbled Circuits & Stateful TEE & GC, OT & IO, IR & IO, exe & \ding{55} & \ding{55} & HBC parties & SIM & {\fontsize{4.5}{5.5} \selectfont Baseline : GC} & {\fontsize{4.5}{5.5} \selectfont Dijkstra: 1.01-200x speedup} \\ \hdashline[0.1pt/0.6pt]
	{[KGM \cite{kaptchuk2017giving}, NDSS, 2019]}                & SOC    & General & Stateless TEE + PL & PRF, DAE, Com & IO, state & IO, exe, state & \ding{55} & \ding{51} & Malicious Server & SIM & \multicolumn{2}{c}{\ding{55}} \\ \hdashline[0.1pt/0.6pt]
	{[Kosto \cite{dang2019towards}, ESORICS, 2019]} & SOC & General & Stateful TEE & PRF, AE, Com & IO & IO, exe & \ding{51} & \ding{55} & Malicious Parties & \ding{55} & {\fontsize{4.5}{5.5} \selectfont Baseline: Plaintext Model} & {\fontsize{4.5}{5.5} \selectfont \makecell[c]{mcf: 2x slowdown \\ deepsjeng: 2.3x slowdown \\ leela : 2.1x slowdown \\ xz : 1.2x slowdown\\ exchange: 0.4x slowdown}} \\ \hdashline[0.1pt/0.6pt]
	{[ObliDC \cite{wu2019oblidc}, AsiaCCS, 2019]}                & SDC, n & General & Stateful TEE + OP & Sig, AE, PKE & IO, pattern, IR & \makecell[c]{IO, exe,\\Completeness} & \ding{55} & \ding{51} & \makecell[c]{Malicious Adaptive\\Servers\\ HBC client} & UC & {\fontsize{4.5}{5.5} \selectfont \makecell[c]{Baseline: Plaintext Model\\ Datasize: 2GB}} & {\fontsize{4.5}{5.5} \selectfont \makecell[c]{WordCount: 0.8-1.2x slowdown \\  RandomWriter: 0.3-1.42x slowdown}} \\ \hdashline[0.1pt/0.6pt]
	{[LucidTEE \cite{sinha2019luciditee}, Preprint, 2019]}       & SMC, n & General & Stateless TEE + PL & \makecell[c]{PKE, Sig,\\AE, Com} & IO, state & \makecell[c]{IO, exe,\\state, policy} & \ding{51} & \ding{51} & Malicious Parties & UC & {\fontsize{4.5}{5.5} \selectfont \makecell[c]{No baseline\\ Datasize: 30MB}} & {\fontsize{4.5}{5.5} \selectfont PSI: 10s} \\ \hdashline[0.1pt/0.6pt]
	{[WNS22\tnote{$\ddagger$} \cite{wu2022hybrid}, NDSS, 2022]}                    & SMC, n & SELECT-JOIN & Stateful TEE & \makecell[c]{PRF, HE,\\BF, CHF} & IO, pattern & IO, exe & \ding{55} & \ding{51} & Malicious Parties & UC & {\fontsize{4.5}{5.5} \selectfont \makecell[c]{Baseline: ABY\\ Function: Select-join}} & {\fontsize{4.5}{5.5} \selectfont Select-join: 2-18kx speedup} \\ \bottomrule
  \end{tabular*}

\begin{tablenotes}[para]
  \scriptsize
  \item \textbf{GC}: Garbled Circuits, \textbf{KE}: Key Exchange, \textbf{SDC,}$n$: Secure Distributed Computation with $n$ Slave Nodes, \textbf{SOC}: Secure Outsourced Computation, \textbf{SMC,}$n$: Secure Multiparty Computation with $n$ Parties, \textbf{AE}: Authenticated Encryption, \textbf{SIM}: Simulate-based Proof, \textbf{HD}: Hamming Distance, \textbf{PSI}: Private Set Intersection, \textbf{Sig}: Signature Scheme, \textbf{PKE}: Public key encryption, \textbf{NIZK}: Non-interactive Zero-knowledge Proof, \textbf{OWF}: One Way Function, \textbf{Com}: Commitment Scheme, \textbf{PRF}: Pseudo-random Functions, \textbf{DAE}: Deterministic Authenticated Ancryption, \textbf{HE}: Homomorphic Encryption, \textbf{BF}: Bloom Filter, \textbf{CHF}: Cuckoo Filter, \textbf{IR}: Intermediate Results, \textbf{PRP}: Pseudo-Random Permutation, \textbf{SecMix}: Secure Mixer, \textbf{OT}: Oblivious Transfer; \\
  \item[$\dagger$] https://github.com/FICS/smcsgx, \item[$\ddagger$] https://github.com/HybrTC/HybrTC.git;
\end{tablenotes}
  \end{threeparttable}\label{tbl:comp_sc}
  \end{table}
\end{landscape}

\subsection{Comparisons} \label{subs:smc_comparisons}

A comprehensive comparison is made in Table \ref{tbl:comp_sc} to analyze existing secure computation protocols with regard to the setting, methodology, security features, and performance. Since TEE provides basic confidentiality and integrity guarantees, most of the existing secure computation protocols are against malicious parties. 
Only LucidiTEE  \cite{sinha2019luciditee} achieves fairness and freshness simultaneously; however, it only works on a very small-scale data set (around 30MB).
Due to the different settings and experimental configurations, we are unable to give an explicit performance comparison among the referenced works in Table  \ref{tbl:comp_sc}.  Instead, we present the performance comparison between a protocol and its baseline protocol.
For example, WNS22 \cite{wu2022hybrid} improves the performance by about 2-18k times compared to its baseline ABY \cite{demmler2015aby} for the select-join function.
\section{Secure Machine Learning using TEEs} \label{sec:machine_learning}
This section turns to review and compare TEE-based protocols towards secure machine learning. Following the frameworks presented in Sec. \ref{sec:criteria},
the protocols are classified into secure outsourced learning, secure distributed learning, and secure multiparty learning.
In secure outsourced learning, we consider two types of computational tasks: training and inference.
Training refers to using training data sets and a machine learning algorithm to generate a trained machine learning model. Inference refers to using a trained machine learning model to make a prediction for a given data item. Secure training and inference are also called privacy-preserving Machine Learning as a Service (MLaaS) \cite{hesamifard2018privacy}. We found that approaches in secure training can also apply to secure inference, therefore we do not consider secure inference in distributed and multiparty learning. Instead, we only consider secure inference under the outsourced learning setting. 

\subsection{Secure Outsourced Learning} \label{subs:caio}

\subsubsection{Problem Statement} 

We review secure outsourced learning from two aspects: secure outsourced training (SOT) and secure outsourced inference (SOI).
~SOT enables a resource-limited client to outsource a training task $(\Phi, \enc{\mathcal{X}})$ to a single cloud server with powerful resources and generate a trained model $M$. Note that $\mathcal{X} =\{(x_i, y_i)\}_{i=1}^{N}$ is an ensemble of data item (e.g., $x_i$) with its label (e.g., $y_i$).
To protect the privacy of training data and model, the cloud server performs training operations $\Phi$ on encrypted data sets $\enc{\mathcal{X}}$ and outputs $(\enc{M}, \Gamma)$, where $\enc{M}$ is an encrypted trained model, and $\Gamma$ is a proof of integrity assurance. The \textit{correctness} of SOT means that the final result $M = \Phi(\mathcal{X})$. In other words, the result of SOT is equivalent to the one trained locally. \textit{Privacy} means that the cloud server cannot learn anything sensitive information about $\mathcal{X}$ (input privacy) or $M$ (model privacy, known as output privacy in SOC). The \textit{integrity} enforces the cloud server performing $\Phi(\enc{\mathcal{X}})$ honestly by generating a proof $\Gamma$ to prove the integrity of execution.

\begin{figure}[ht]
  \centering
  \subcaptionbox{Secure Outsourced Training\label{subfig:SOT}}{%
    \includegraphics[width=0.48\textwidth]{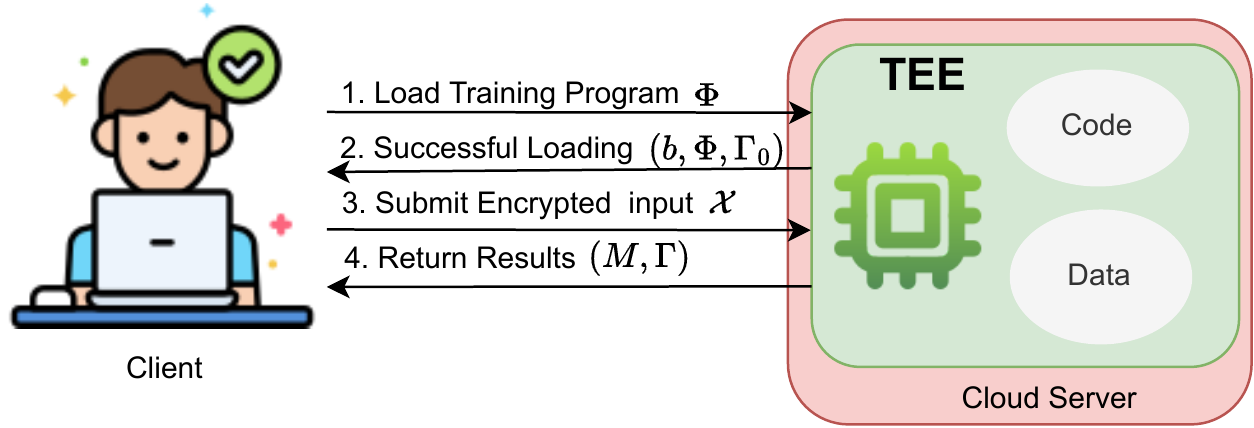}
  }
  \subcaptionbox{Secure Outsourced Inference\label{subfig:SOI}}{%
    \includegraphics[width=0.48\textwidth]{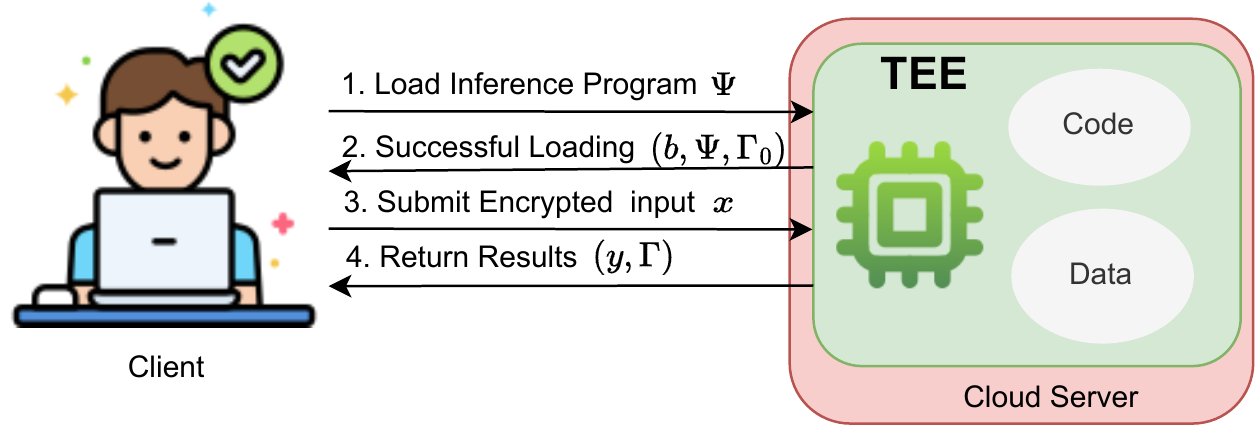}
  }
  \caption{Secure Outsourced Learning}\label{fig:caio} 
  \Description{Secure outsourced training and inference.}
\end{figure}

While cryptographic primitives (e.g., homomorphic encryption, secret sharing, secure multi-party computation) can support secure training \cite{boulemtafes2020review}, they introduce high computation and communication overhead \cite{duy2021confidential}. TEE-based SOT has gain its popularity in recent years due to its appealing performance advantages. As shown in Fig. \ref{subfig:SOT}, we provide a general workflow of SOT to explain the interactions between a client and a TEE-equipped cloud server. The client builds a secure channel between the enclave by remote attestation, after which four steps are followed. (1) The client loads the program $\Phi$ to the enclave via a secure channel by remote attestation. (2) The enclave responds with a triple $(b, \Phi, \Gamma_0)$, in which $\Gamma_0$ proves the program is loaded correctly ($b=1$); otherwise, $b=0$ and $\Gamma_0 = \bot$. (3) The client submits an encrypted data ensemble $\enc{\mathcal{X}}$ to the enclave. (4) The enclave cooperates with the untrusted cloud server to generate the final results $(\enc{M}, \Gamma)$, where $M = \Phi(\mathcal{X})$ and $\Gamma$ proves the program is executed honestly. The enclave returns the final results to the client.

In SOI, a single cloud server holds the trained model $M$ and allows a resource-limited client to submit an inference task $(\Psi, \enc{x})$, and then responds the client with an inference output $y$. Note that here $x$ is just a single data item and the output $y$ is the prediction result.
To protect the privacy of $x$, the cloud server performs inference operations $\Psi$ on encrypted inference data $\enc{x}$. It outputs $(\enc{y}, \Gamma)$, where $\enc{y}$ is an encrypted inference result, and $\Gamma$ is a proof of integrity for operations. The \textit{correctness} of SOI means that the final inference result $y = \Psi(x, M)$. In other words, the result of SOI is equal to the one that finishes the inference task locally. The SOI \textit{privacy} means that the cloud server cannot learn anything sensitive information about $x$ (input privacy) or $y$ (output privacy). In case the cloud server has no ownership of the model $M$, the encrypted model $\enc{M}$ is fed to server to achieve the model privacy.
The \textit{integrity} enforces the cloud server to perform $\Psi(\enc{x}, M)$ honestly and generate the integrity proof $\Gamma$.

Fig. \ref{subfig:SOI} shows a general workflow of SOI and the interactions between a client and a single cloud server that equips TEE and holds a machine learning model $M$. The client builds a secure channel between the enclave by remote attestation, after which five steps are followed. (1) The client loads the program $\Psi$ to the enclave via a secure channel. (2) The enclave responds with a tuple $(b, \Psi, \Gamma_0)$, in which $\Gamma_0$ proves the program is loaded successfully ($b=1$); otherwise, $b=0$ and $\Gamma_0 = \bot$. (3) The client submits an encrypted input $\enc{x}$ to the enclave. (4) The enclave cooperates with the cloud server to generate the final results $(\enc{y}, \Gamma)$, in which $\enc{y} = \Psi(\enc{x}, M)$ and $\Gamma$ proves that the program is executed honestly. The enclave returns the final results to the client.

\subsubsection{Research status of SOT}

In \cite{ohrimenko2016oblivious}, Ohrimenko et al. combined SGX and data-oblivious primitives to design data-oblivious machine learning training algorithms, such as support vector machines, matrix factorization, neural networks, decision trees, and k-means clustering. Experimental evaluations show that its performance outperforms the  MPC-based solutions. However, the approach results in a large TCB, which limits its use to support large machine learning models. To reduce TCB and enable the TensorFlow framework, Kunkel et al. \cite{kunkel2019tensorscone} presented the \textsc{TensorScone} framework to support secure machine learning computations on untrusted infrastructure. \textsc{TensorScone} can achieve more than 80\% precision compared with GPU-only baseline training. In \cite{yuhala2021plinius}, Peterson et al. proposed Plinius that combined the features between persistent memory (PM) and Intel SGX enclave. Plinius achieves  3.2-3.7$\times$ performance improvement for saving and restoring models on real PM hardware, respectively. 
Besides, GPU-based TEEs \cite{volos2018graviton} were introduced to accelerate secure training (e.g., Telekine \cite{hunt2020telekine}).
To enable parallel GPU acceleration, the proposed AsymML in \cite{niu2021asymml} partitions a target machine learning model into trusted and untrusted parts. Thus, AsymML can provide more than 5$\times$ speedup than pure TEE-based training. Besides, model partition is also a popular technology to accelerate the TEE-based training \cite{hashemi2021darknight,yuhala2021plinius,asvadishirehjini2022ginn}.

To enable realistic integrity-preserving deep neural network (DNN) model training for heavy workloads, Asvadishirehjini et al. \cite{asvadishirehjini2022ginn} presented GINN that combined random verification of selected computation steps with systematic adjustments of DNN hyperparameters to limit the attacker’s ability to shift the model parameters arbitrarily. GINN can achieve high integrity and 2-20$\times$ performance improvement over a pure TEE-based training method. In \cite{hashemi2021darknight}, to optimize the performance, Hashemi et al. presented DarKnight framework that uses TEE to provide privacy and integrity verification and GPUs to perform the bulk of linear algebraic computation. 
DarKnight \cite{hashemi2021darknight} proposed a customized data encoding strategy based on matrix masking, which is adopted to create input obfuscation within the TEE. DarKnight achieves an average of 6.5$\times$ training speedup compared with the pure TEE-based solutions.
Goten \cite{ng2021goten} adopted a dynamic quantization scheme to cater for the fluctuation in weight during training to optimize performance. It shows a 6.84-132.64$\times$ speedup than a pure TEE-based solution and the latest secure multi-server cryptographic solution \cite{wagh2021falcon}. In \cite{ozga2021perun}, Ozga et al. presented \textsc{Perun} that considers multi-stakeholder collaboratively generating a trained model where the stakeholders consist of the training data owner, training code owner, model owner, and inference code owner. Each entity outsources the computation to a cloud server and provides a service to other entities. \textsc{Perun} gives a 161-1,560$\times$ performance improvement compared with a pure TEE-based method.

\subsubsection{Research status of SOI}

Due to the similarity of computation between training and inference, many existing works \cite{hunt2020telekine,niu2021asymml,hashemi2021darknight,ng2021goten,yuhala2021plinius,ozga2021perun} support SOT and SOI simultaneously. In the following, we review other works focusing on SOI. In \cite{grover2018privado}, Grover et al. first pointed out the inference of DNN based on Intel SGX enclaves faced with access pattern based attacks, so this work designed input-oblivious \textsc{Privado} framework to address the problem and support secure and integrated inference services. \textsc{Privado} with Torch framework incurs an average 17.18\% overhead on 11 different DNN models. The work \cite{gu2018securing} formulated the information exposure problem as a reconstruction privacy attack for a cloud-based inference service and quantified the adversary's capabilities with different attack strategies. DeepEnclave was proposed in \cite{gu2018securing} to partition a deep learning model into a FrontNet and a BackNet, where the FrontNet is enforced to do enclaved execution, and the BackNet runs out of secure enclaves. DeepEnclave causes 1.64-2.54$\times$ overhead due to the enclaved execution of the FrontNet compared to GPU execution of the FrontNet. To optimize the performance of inference, Florian et al. proposed \textsc{Slalom} \cite{florian2019slalom} that partitioned DNN computations into trusted and untrusted parts. All the computation of linear layers of a DNN model are executed in an enclave, while other computation of the DNN model are loaded to an untrusted GPU for acceleration. Although \textsc{Slalom} adopts some cryptographic primitives to support verifiable and private inference, it achieves 4-20$\times$ performance improvement compared with a pure TEE approach. 

To break through the memory limitation of the enclave and accelerate an inference, Origami \cite{narra2019privacy} combines enclave, cryptographic tools, and GPU/CPU acceleration. The enclave adds noise to obfuscate inference data and sends the obfuscated data to an untrusted GPU/CPU for computation acceleration. Origami achieves 11$\times$ performance improvement than a pure SGX approach, and 15.1$\times$ performance improvement compared to \textsc{Slalom} \cite{florian2019slalom}. Aiming to address the memory limitation and page swapping of SGX enclave, Lee et al. \cite{lee2019occlumency} carefully developed on-demand weights loading, memory-efficient inference, and parallel processing pipelines in their proposed \textsc{Occlumency}, which gives a 3.6$\times$ speedup compared to a pure TEE-based method and 72\% latency overhead compared to a pure GPU approach. In \cite{schlogl2020ennclave}, Alexander et al. presented the \textsc{eNNclave} tool-chain to cut TensorFlow models at any layers and split them into public and enclave layers, where GPU performs public layers for acceleration. In contrast, the enclave executes private layers for security and integrity. Accuracy results of \textsc{eNNclave} are close to a pure GPU method. Aiming to solve the large memory allocation and low memory re-usability problem in deep learning systems, \cite{kim2020vessels} presented \textsc{Vessels} that optimized the memory management of SGX. However, in contrast to a pure GPU-based inference method, \textsc{Vessels} results in about 3-20$\times$ inference latency.

To provide dependable and timely inference services, Xiang et al. \cite{xiang2021aegisdnn} proposed AegisDNN that adopted a dynamic-programming algorithm based on the layer-wise DNN time and Silent Data Corruption profiling mechanism to find a layer protection configuration for each inference task. AegisDNN supports Caffe, PyTorch, and TensorFlow frameworks and achieves up to 88.8\% less failure rates than the pure GPU-based approach and 99.9\% shorter relative response time than the pure enclave-based approach. Different from the previous works, \cite{park2022fairness} focused on the fairness of inference for the machine learning model with the enclave.

\subsection{Secure Distributed Learning} \label{subs:caid}


\subsubsection{Problem Statement}
Secure distributed training (SDT) enables $n$ ($n \geq 2$) slave nodes to jointly perform a training task $(\Phi, \mathcal{X}_1, \cdots, \mathcal{X}_n)$ with the assistance of a master node and generate a trained model $M$, where $\mathcal{X}_i$ is the $i$-th party's training data. Particularly, $\Phi$ consists of slaves' local training operation $\Phi_i$ and master's aggregation operation $\Pi$ such that
\begin{equation}
\Phi(x) = \Pi\left(\Phi_1(\mathcal{X}_1),\Phi_2(\mathcal{X}_2), \cdots, \Phi_N(\mathcal{X}_N)\right).
\end{equation}
Note that each slave node performs a local training $\Phi_i$ and outputs intermediate result $(\enc{m_i}, \Gamma_i)$, where $\enc{m_i}$ is an encrypted intermediate model (also called local model), and $\Gamma_i$ proves the local training is executed honestly. After that, master node aggregates the local models into one global model.
The \textit{privacy} means that master node and slave nodes fail to learn any sensitive information about $\mathcal{X}$ (input privacy) or $M$ (output privacy). The \textit{integrity} enforces the master node and the slave nodes to honestly perform $\Pi$ and $\Phi_i$ and generate the respective proofs.

\begin{figure}[h]
  \centering
  \includegraphics[width=\linewidth]{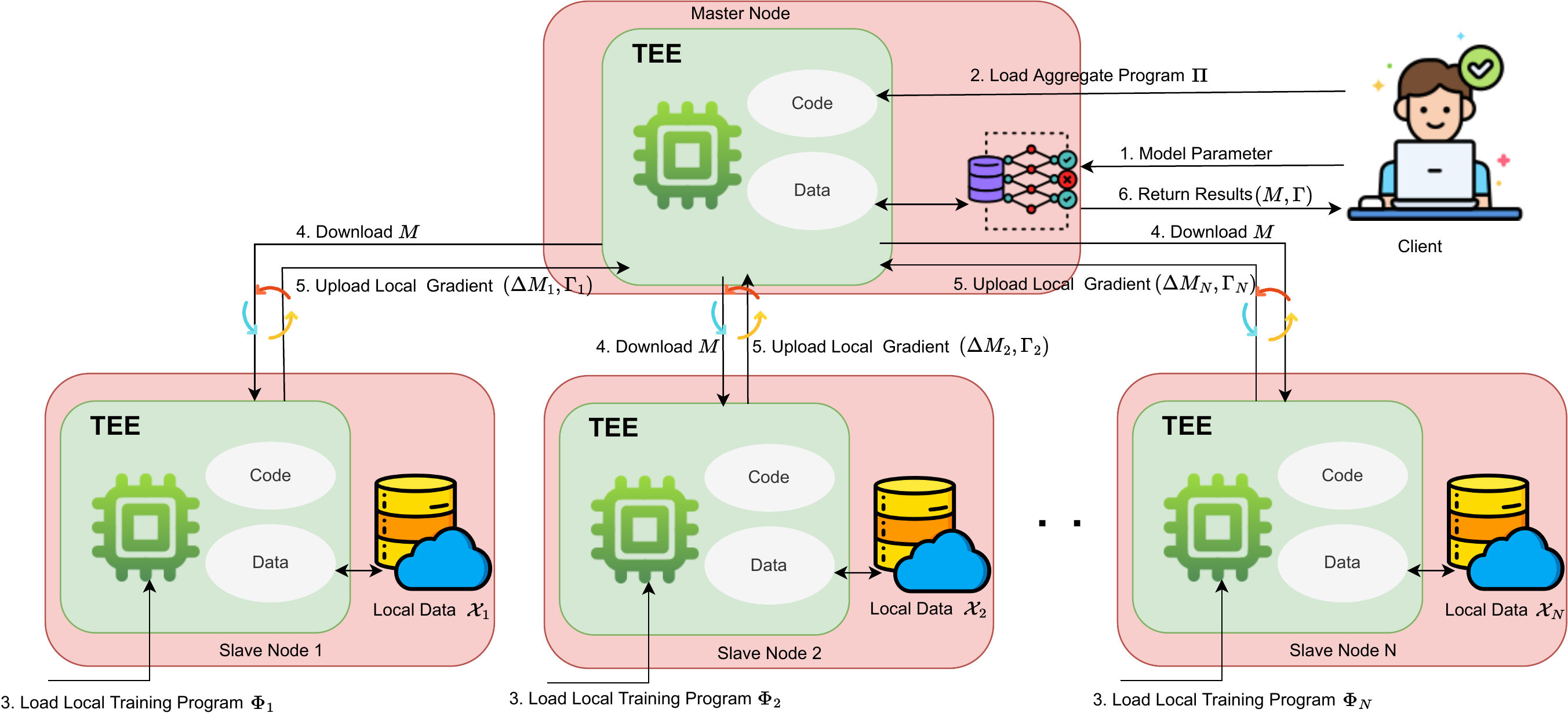}
  \caption{Secure Distributed Training}\label{fig:sdt}
  \Description{Secure Distributed Learning}
\end{figure}

As depicted in Fig. \ref{fig:sdt}, we present a general workflow of TEE-based SDT to explain the interactions among a client, multiple TEE-equipped parties, and a TEE-equipped master node. The client builds a secure channel between the master node's enclave via remote attestation, after which six steps are followed.
(1) The client uploads the model parameter to the master node.
(2) The client ensures the aggregation program $\Pi$ is loaded to the master node's enclave successfully and (3) ensures the local training algorithm $\Phi_i$ is loaded into slaves' enclave successfully.
(4) The slave nodes download the initial global model $M$ and then generate a model gradient $\Delta M$.
(5) The slave nodes submits the gradient $\Delta M$ to the master node. Note that the steps (4) and (5) are repeated until a convergent model is generated or the maximum number of iterations is reached.
(6) Finally the master node returns the final global model to the client.

Note that Federated Learning (FL) also belongs to the distributed framework. FL serving as a special scenes of SDT has recently attracted much attention from the research and industrial community. The input data in SDT can either be encrypted or unencrypted. However in FL, each slave has its own training data set and a local model over local unencrypted data. The \textit{privacy} for FL means that master node fails to learn any sensitive information about $\mathcal{X}$ (input privacy) or $M$ (output privacy). The \textit{integrity} enforces the master node and slave nodes performing $\Pi$ and $\Phi_i$ honestly and generating the proof accordingly, respectively.

\subsubsection{Research Status}

In \cite{hunt2018chiron}, the proposed Chiron adopted SGX and the sandbox Ryoan \cite{hunt2016ryoan} to perform model training and protect the privacy of both training data and the trained model, where Ryoan sandbox was used to prevent leaking the training data outside the enclave. Particularly, Chiron supports distributed training with multiple enclaves. Compared with training deep neural networks on CPU-only, Chiron increases about 4\%-20\% training cost. The work \cite{zhang2021citadel} proposed Citadel that allows parties (or say data owners) holding training data and an aggregation server (or say aggregator owner) equipped with enclaves to protect their input privacy. When multiple enclaves are deployed, Citadel achieves a less than 1.73$\times$ performance slowdown. 
In contrast to other solutions, \cite{huang2021starfl} proposed StarFL that combines multiple techniques, such as TEE, MPC, and satellites (who are responsible for distributing keys). When the last four layers are inside enclave execution, the running time of StarFL is no more than 5\% higher than a pure CPU-based training approach.
Brito et al. presented \textsc{Soteria} \cite{brito2021soteria}, a distributed privacy-preserving machine learning (ML) system based on TEEs. \textsc{Soteria} made performance trade-offs for the distributed Apache Spark framework and its ML library. As \textsc{Soteria} supports executing non-sensitive operations outside an enclave, it improves the runtime performance of distinct ML algorithms by up to 1.7$\times$ compared to a pure SGX-spark method.

To support unmodified TensorFlow applications, Quoc et al. presented a distributed secure machine learning framework (\textsc{secureTF}) based on TensorFlow for the untrusted cloud infrastructure \cite{quoc2020securetf}. In contrast to Chiron \cite{hunt2018chiron}, \textsc{secureTF} support SDT and SDI simultaneously and enjoys the advantage of unmodified TensorFlow applications. But it endures roughly 14$\times$ training overhead compared to a pure GPU approach.

\noindent\textbf{Federated Learning.} SecureFL \cite{kuznetsov2021securefl} equips slave nodes and the master node of FL with TEEs. However, due to the partial  in-enclave training, SecureFL results in 1.6\%-23.6\% overhead compared with a pure CPU-based training. Aiming to limit privacy leakages in FL, Mo et al. presented a privacy-preserving federated learning (PPFL) framework \cite{mo2020darknetz} for mobile systems using TEE. PPFL adopted TEE to train each layer of the model until its convergence by leveraging the greedy layer-wise. Thanks to the TEE, PPFL can defend against data reconstruction, property inference, and membership inference attacks. In contrast to the standard FL without privacy protection, PPFL introduces about 1.3s-1.5$\times$ training overhead.

To improve the efficiency, TrustFL proposed in \cite{zhang2020enabling} suggested using GPU to train a local model for each party and then verifying the correctness in an enclave. TrustFL utilized a "commit-and-prove" MHT-based commitment scheme to verify the integrity of the training algorithm and a "dynamic-yet-deterministic" data selection strategy for fresh training data. Thanks to GPU-based training, TrustFL provides 1-2 orders of magnitude speedup than a pure TEE-based FL method.  Aiming to defend against Byzantine failures in FL and enable secure model aggregation, Zhao et al. proposed SEAR based on TEE \cite{zhao2021SEAR}. The key idea of SEAR is to build a secure channel between a party and the aggregation server by remote attestation, and encrypted local models are aggregated in the aggregation server's enclave. SEAR achieves a 4$\times$-6$\times$ performance improvement compared to the existing secure aggregation framework. ShuffleFL \cite{zhang2021shufflefl} is proposed to protect the privacy of model gradients and defend against side-channel attacks. It combined random group structure and intra-group gradient segment aggregation mechanisms to prevent the exploitation of side-channel attacks. Like TrustFL, ShuffleFL also uses enclave to perform model aggregation.

\subsection{Secure Multiparty Learning} \label{subs:caim}
This section turns to review and compare TEE-based protocols towards secure multi-party training.

\subsubsection{Problem Statement}

Secure multi-party training (SMT) enables $n$ ($n \geq 2$) parties holding data jointly performing a training task $(\Phi, \mathcal{X}_1, \cdots, \mathcal{X}_N)$ and generating a trained model $M$, where $\mathcal{X}_i$ is the $i$-th party's training data.
The \textit{correctness} of SMT means that the trained model is equal to the model that was trained on an honest party over the same data set. The \textit{privacy} means that any party fails to learn any sensitive information about other parties' data. The \textit{integrity} enforces each party performing $\Psi_i(x_i)$ honestly and generating the operation integrity proof  $\Gamma_i$.

\begin{figure}[h]
  \centering
  \includegraphics[width=0.7\linewidth]{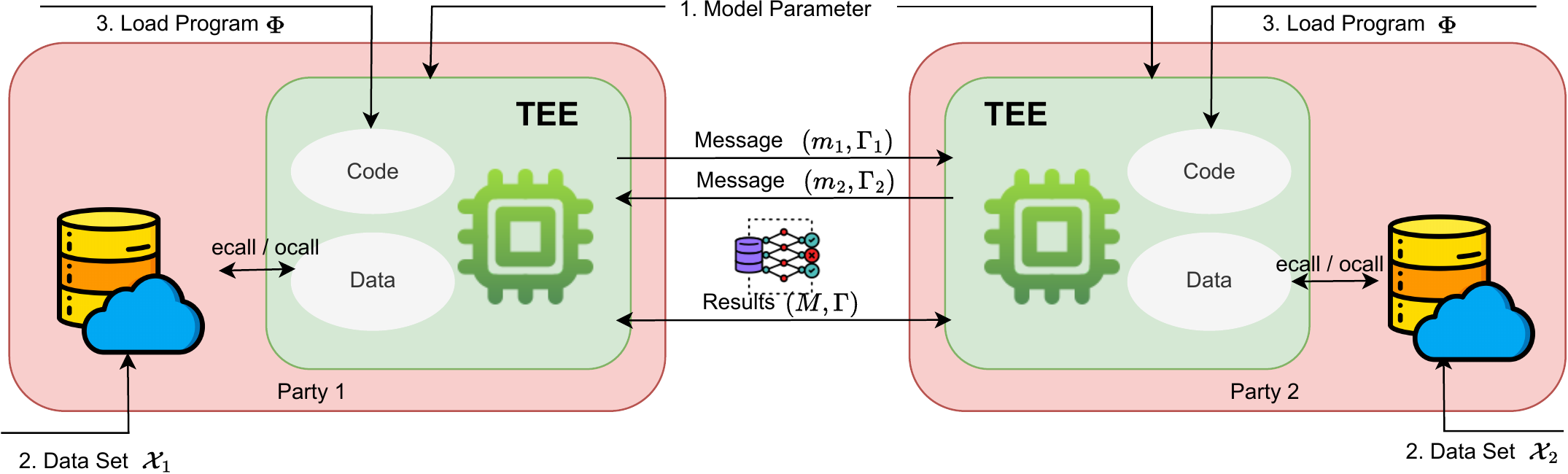}
  \caption{Secure Multiparty Training}\label{fig:SMT}
  \Description{Assessment}
\end{figure}

Fig. \ref{fig:SMT} shows a general workflow of secure two party training between two TEE-equipped parties, which can be easily extended to the multi-party setting. The two party builds a secure channel via remote attestation, after which all messages are communicated over the secure channel. Then three steps are followed.
(1) Both enclave read the model parameter into TEE.
(2) Both parties take their own data $\mathcal{X}_1$ and $\mathcal{X}_2$ as input, respectively.
(3) Both parties ensure the training program $\Phi$ is loaded into the TEEs successfully.
After that the protocol executes interactively between the two parties. In each interaction, a message associated with a proof $(m_i, \Gamma_i)$ is sent to another party. Here $m_i$ is a committed message to ensure the privacy of the execution, and $\Gamma_i$ is employed to prove the message $m_i$ is computed honestly. 
The protocol halts until generating a convergent model and outputs a final trained model $(M, \Gamma)$, where $\Gamma$ proves the protocol is run honestly.

\subsubsection{Research Status}
Although secure multi-party computation (MPC) is successfully applied to secure machine learning \cite{choi2019secure}, only few solutions of multi-party machine learning training are designed without requiring a third party (or master node) to participate in computations. One possible explanation is that the training algorithms faces with complex computations, which affects the performance significantly. 
In \cite{ren2021hybrid}, to reduce the communication overhead of MPC-based methods for privacy-preserving neural networks training, Ren et al. presented an $n$-party training framework for Graph Neural Networks (GNNs) using SGX. During training, $n$ parties hold training data, and SGX shares training data by additive secret sharing and perform training operations on the local enclave. Following that, the party aggregates other $n - 1$ parties' training results to obtain the trained model. The solution of \cite{ren2021hybrid} can output a trained GNN within dozens of seconds.

\begin{landscape}
  \begin{table}[t]
    \caption{Comparisons of Secure Machine Learning}
    \begin{threeparttable}
      \setlength\tabcolsep{1.5pt}
\renewcommand{\arraystretch}{1.3}
\tiny
\begin{tabular*}{\linewidth}{cccccccccccc}
    \toprule
    \multirow{3}*{\scriptsize Schemes} & \multicolumn{2}{c}{\multirow{2}*{\scriptsize Settings}} & \multicolumn{2}{c}{\multirow{2}*{\scriptsize Methodology}} & \multicolumn{5}{c}{\scriptsize Security} &  \multicolumn{2}{c}{\multirow{2}*{\scriptsize Performance}} \\ \cmidrule(lr){6-10} 
                &   &   &   &   & \multicolumn{3}{c}{Security Goals} & \multirow{2}*{\makecell[c]{Adversarial\\Models}} & \multirow{2}*{\makecell[c]{Formal\\proof}} &   & \\ \cmidrule(lr){2-3} \cmidrule(lr){4-5} \cmidrule(lr){6-8} \cmidrule(lr){11-12}  
                & Framework & Function & Technologies & Primitives & Privacy & Integrity & Freshness & & & Configurations & Costs \\ \midrule
    {[OMPML \cite{ohrimenko2016oblivious}, USENIX Security, 2016]}      & SOT & Training & TEE+OP & \makecell[c]{AES-GCM,\\GPO} & IO, exe, pattern & IO, exe & \ding{51} & Malicious Server/Client & \ding{55} &  {\fontsize{4.5}{5.5} \selectfont \makecell[c]{Dataset: MNIST, SUSY, MovieLens, Nursery\\Models: K-Means, CNN, SVM, Matrix factorizatoin\\Baseline: MPC}} & 1.7-2.5x speedup\\ \hdashline[0.1pt/0.6pt]
    {[\textsc{Privado} \cite{grover2018privado}, arXiv, 2018]}          & SOI & Inference & TEE & DO,TLS & IO, exe & IO, exe & \ding{55} & Malicious Server & \ding{55} &  {\fontsize{4.5}{5.5} \selectfont \makecell[c]{Dataset: MNIST, CIFAR10\\Models: MLP, LeNet, VGG19, Wideresnet,\\ ResNet29/50/110, AlexNet, Squeezenet,\\ InceptionV3, DenseNet\\Platform: Intel SGX SDK, torch\\Baseline: Plaintext Model}} & 0.18x slowdown \\ \hdashline[0.1pt/0.6pt]
    {[Chiron \cite{hunt2018chiron}, arXiv, 2018]}                       & SDT & Training & TEE & \makecell[c]{AES-GCM,\\DF, TLS} & IO, exe & IO, exe & \ding{55} & Malicious Server & \ding{55} &  {\fontsize{4.5}{5.5} \selectfont \makecell[c]{Dataset: CIFAR-10, ImageNet\\Models: VGG-9, AlexNet\\Platforms: Intel SGX, Ryoan sandbox\\Baseline: Plaintext Model}} & 0.04-0.2x slowdown \\ \hdashline[0.1pt/0.6pt]
    {[DeepEnclave \cite{gu2018securing}, arXiv, 2018]}                  & SOI & Inference & TEE & AE, TLS & IO, exe & IO, exe & \ding{55} & Malicious Server & \ding{51} &{\fontsize{4.5}{5.5} \selectfont \makecell[c]{Dataset: ImageNet\\Models: Darknet, Extraction, DeseNets\\Baseline: Pliantext Model}} & 1.64-2.54x slowdown \\ \hdashline[0.1pt/0.6pt]
    {[\textsc{Slalom} \cite{florian2019slalom}, ICLR, 2019]}            & SOI & Inference & TEE & \makecell[c]{Frei., SC\\PRNG} & \makecell[c]{IO\\(model, intput)} & exe & \ding{55} & Malicious Server & \ding{51} & {\fontsize{4.5}{5.5} \selectfont \makecell[c]{Dataset: ImageNet\\Models: VGG16, MobileNet\\Platforms: Intel SGX, TensowFlow\\Baseline: Fully in Enclave}} & 4-20x speedup\\ \hdashline[0.1pt/0.6pt]
    {[\textsc{TensorSCONE} \cite{kunkel2019tensorscone}, arXiv, 2019]}  & SOT & Training & TEE & TLS & IO, exe & IO, exe & \ding{51} & Malicious Server & \ding{55} & {\fontsize{4.5}{5.5} \selectfont \makecell[c]{Dataset: CIFAR-10\\Models: InceptionV4\\Platforms: Intel SGX, SCONE, TensorFlow\\Baseline: Plaintext Model}} & 0.8-0.9x slowdown \\ \hdashline[0.1pt/0.6pt]
    {[\textsc{Occlumency} \cite{lee2019occlumency}, MobiCom, 2019]}     & SOI & Inference & TEE & AES, TLS & IO, exe & IO, exe & \ding{55} & Malicious Server & \ding{55} & {\fontsize{4.5}{5.5} \selectfont \makecell[c]{Dataset: ImageNet\\Models: AlexNet, GoogleNet, ResNet50/101/152,\\ VGG16/19, YOLO\\Platforms: Intel SGX, Caffe}} & {\fontsize{5.5}{7} \selectfont \makecell[c]{Fully in Enclave: 3.6x speedup \\Plaintext Model: 0.72x slowdown}} \\ \hdashline[0.1pt/0.6pt]
    {[Origami \cite{narra2019privacy}, arXiv, 2019]}                    & SOI & Inference & TEE & - & IO & - & \ding{55} & Malicious Server & \ding{55} & {\fontsize{4.5}{5.5} \selectfont \makecell[c]{Dataset: ImageNet\\Models: VGG16/19\\Baseline:Slalom}} & 11-15.1x speedup \\ \hdashline[0.1pt/0.6pt]
    {[Telekine \cite{hunt2020telekine}, NSDI, 2020]}                    & SOT, SOI & \makecell[c]{Training\\Inference} & GPU TEE+OP & \makecell[c]{AES-GCM,\\MAC, LTS} & IO, pattern & IO, exe & \ding{51} & Malicious Server & \ding{55} & {\fontsize{4.5}{5.5} \selectfont \makecell[c]{Dataset: ImageNet\\Models: ResNet, InceptionV3, DenseNet\\Platforms: AMD's ROCm 1.8, MXNet\\Baseline: Plaintext Model}} & {\fontsize{5.5}{5.5} \selectfont \makecell[c]{Training: 1.08-1.23x slowdown\\Inference: 1.0-10x slowdown}} \\ \hdashline[0.1pt/0.6pt]
    {[eNNclave \cite{schlogl2020ennclave}, AISec, 2020]}               & SOI & Inference & TEE & AES-GCM & IO & IO & \ding{55} & Malicious Server/Client & \ding{55} &  {\fontsize{4.5}{5.5} \selectfont \makecell[c]{Dataset: MIT67, Flowers, Amazon review data\\Models: VGG-16, VGG-19, CNN\\Baseline: Plaintext Model}} & 17x-65x slowdown \\ \hdashline[0.1pt/0.6pt]
    {[\textsc{Vessels} \cite{kim2020vessels}, SoCC, 2020]}              & SOI & Inference & TEE & - & IO, exe & IO, exe & \ding{55} & Malicious Server & \ding{55} &  {\fontsize{4.5}{5.5} \selectfont \makecell[c]{Dataset: ImageNet\\Models: AlexNet, ResNet101, ResNet152, DenseNet201,\\ ResNext152, DarkNet53, InceptionV3, VGG16, YoloV3\\Baseline: Plaintext Model}} & Inference: 5.01-20.9x slowdown\\ \hdashline[0.1pt/0.6pt]
    {[ShadowNet \cite{sun2020shadownet}, arXiv, 2020]}                  & SDT & \makecell[c]{Training\\Inference} & TEE & - & Model & - & \ding{55} & Malicious Server/Client & \ding{51} &  {\fontsize{4.5}{5.5} \selectfont \makecell[c]{Dataset: ImageNet, CIFAR-10\\Models: MobileNets, ResNet-44, AlexNet, MiniVGG \\Platforms: TensowFlow Lite\\Baseline: Fully in Enclave}} & 0.61-2.21 speedup \\ \hdashline[0.1pt/0.6pt]
    {[TrustFL \cite{zhang2020enabling}, INFOCOM, 2020]}                 & SDT & Training & TEE & PRF, HMAC & IO, exe & IO, exe & \ding{55} & \makecell[c]{Semi-honest Server\\Dishonest Participant} & \ding{55} &  {\fontsize{4.5}{5.5} \selectfont \makecell[c]{Dataset: CIFAR10, ImageNet\\Models: VGG16/19, ResNet\\Baseline: Fully in Enclave}} & 10-100x speedup\\ \hdashline[0.1pt/0.6pt]
    {[DarkneTZ \cite{mo2020darknetz}, MobiSys, 2020]}                   & SDT & \makecell[c]{Training\\Inference} & TEE & AES-GCM & IO, exe & IO, exe & \ding{55} & Malicious Client & \ding{55} & {\fontsize{4.5}{5.5} \selectfont \makecell[c]{Dataset: CIFAR100, ImageNet Tiny\\Models: AlexNet, VGG7\\Platforms: ARM TrustZone\\Baseline: Plaintext Model}} &  0.03-0.1x slowdown \\ \bottomrule
\end{tabular*}

\begin{tablenotes}[para]
    \scriptsize
    \item \textbf{GPO}: General-purpose Oblivious, \textbf{DO}: Data Oblivious, \textbf{Frei.}: Freivalds' algorithm, \textbf{SC}: Stream ciphers, \textbf{PRNG}: Pseudo Random Number Generator, \textbf{AE}: Authenticated Encryption, \textbf{MPC}: Multiple Party Computation
\end{tablenotes}
    \end{threeparttable}\label{tbl:comp_ppml}
    \end{table}
  \end{landscape}

  \newpage

  \begin{landscape}
    \begin{table}[t]
      \begin{threeparttable}
        \setlength\tabcolsep{1.3pt}
\renewcommand{\arraystretch}{1.3}
\tiny
\begin{tabular*}{\linewidth}{cccccccccccc}
    \toprule
    \multirow{3}*{\scriptsize Schemes} & \multicolumn{2}{c}{\multirow{2}*{\scriptsize Settings}} & \multicolumn{2}{c}{\multirow{2}*{\scriptsize Methodology}} & \multicolumn{5}{c}{\scriptsize Security} &  \multicolumn{2}{c}{\multirow{2}*{\scriptsize Performance}} \\ \cmidrule(lr){6-10} 
                &   &   &   &   & \multicolumn{3}{c}{Security Goals} & \multirow{2}*{\makecell[c]{Adversarial\\Models}} & \multirow{2}*{\makecell[c]{Formal\\proof}} &   & \\ \cmidrule(lr){2-3} \cmidrule(lr){4-5} \cmidrule(lr){6-8} \cmidrule(lr){11-12}  
                & Framework & Function & Technologies & Primitives & Privacy & Integrity & Freshness & & & Configurations & Costs \\ \midrule
    {[\textsc{secureTF} \cite{quoc2020securetf}, Middleware, 2020]}     & SDT & \makecell[c]{Training\\Inference} & TEE & ECDH, TLS & IO, exe & IO, exe & \ding{55} & Malicious Server & \ding{55} & {\fontsize{4.5}{5.5} \selectfont \makecell[c]{Dataset: CIFAR10, MNIST\\Models: Inception v3/v4, DeseNet\\Platforms: Intel SGX, TensowFlow\\Baseline: Plaintext Model}} & 14x slowdown \\ \hdashline[0.1pt/0.6pt]
    {[AegisDNN \cite{xiang2021aegisdnn}, RTSS, 2021]}                   & SOI & Inference & TEE & - & IO, exe & IO, exe & \ding{55} & Malicious Server & \ding{55} & {\fontsize{4.5}{5.5} \selectfont \makecell[c]{Dataset: -\\Models: lenet, Alexnet, ResNet-18, Pilotnet\\Platforms: Intel SGX, TenseFlow, Caffe, Pytorch\\Baseline: Pliantext Model}} & 0.8-1.2 slowdown \\ \hdashline[0.1pt/0.6pt]
    {[AsymML \cite{niu2021asymml}, arXiv, 2021]}                        & SOT, SOI & \makecell[c]{Training\\Inference} & TEE & - & IO (model, intput) & - & \ding{55} & Malicious Server & \ding{55} & {\fontsize{4.5}{5.5} \selectfont \makecell[c]{Dataset: ImageNet, CIFAR-10\\Models: VGG16/19, ResNet18/34\\Baseline: Fully in Enclave}} & {\fontsize{5.5}{5.5} \selectfont \makecell[c]{Training: 11.2x speedup \\Inference: 7.6x speedup}} \\ \hdashline[0.1pt/0.6pt]
    {[SEAR \cite{zhao2021SEAR}, TDSC, 2021]}                            & SDT & Training & TEE & \makecell[c]{AES-GCM,\\Elliptic Curve} & IO, exe & IO, exe & \ding{55} & \makecell[c]{Semi-honest\\Server/Client} & \ding{55} & {\fontsize{4.5}{5.5} \selectfont \makecell[c]{Dataset: MNIST, CIFAR-10\\Models: CNN, ResNet56\\Baseline: ASS-based solution}} & 4-6x speedup \\ \hdashline[0.1pt/0.6pt]
    {[SecDeep \cite{liu2021secdeep}, IoTDI, 2021]}                      & SDT & \makecell[c]{Training\\Inference} & TEE & AES-CTR, MD5 & IO, exe & IO, exe & \ding{55} & Malicious Server & \ding{55} & {\fontsize{4.5}{5.5} \selectfont \makecell[c]{Dataset: MNIST, CIFAR-10\\Models: SqueezeNet, MobileNet V1/V2, GoogleNet,\\ Yolo Tiny, ResNet50, Inception BN\\Platforms: HiKey960 board,\\ARM TrustZone \\Baseline: Pliantext Model}} & 16-172x slowdown \\ \hdashline[0.1pt/0.6pt]
    {[Citadel \cite{zhang2021citadel}, SoCC, 2021]}                       & SDT & Training  & TEE & \makecell[c]{AES-CBC-256\\ZSM, TSH} & IO, exe & IO, exe & \ding{55} & \makecell[c]{Semi-honest\\Server/Client} & \ding{55} & {\fontsize{4.5}{5.5} \selectfont \makecell[c]{Dataset: MNIST, SMS messiage\\Models: AlexNetS, AlexNetL, SpamNet, MNIST\\Platforms: Intel SGX, TensowFlow\\Baseline: Fully in Enclave}} & Training: 1.09-1.73x slowdown \\ \hdashline[0.1pt/0.6pt]
    {[Mlcapsule \cite{hanzlik2021mlcapsule}, CVPR, 2021]}               & SDT & \makecell[c]{Training\\Inference} & TEE & - & IO, exe & IO, exe & \ding{55} & Malicious Client & \ding{55} & {\fontsize{4.5}{5.5} \selectfont \makecell[c]{Dataset: CIFAR100\\Models: VGG-16\\Baseline: Plaintext Model}} & Inference: 1.2-3x overhead \\ \hdashline[0.1pt/0.6pt]
    {[StarFL \cite{huang2021starfl}, ACM TIST, 2021]}                   & SDT & Trainting & TEE & AES-GCM & IO, exe & IO, exe & \ding{55} & Malicious Client & \ding{55} & {\fontsize{4.5}{5.5} \selectfont \makecell[c]{Dataset: CIFAR100, ImageNet Tiny\\Models: AlexNet, VGG7\\Platforms: ARM TrustZone\\Baseline: Plaintext Model}} & less than 5\% slowdown \\ \hdashline[0.1pt/0.6pt]
    {[Model Protection \cite{hou2021model}, TDSC, 2021]}                & SDT & \makecell[c]{Training\\Inference} & TEE & - & Model Privacy & IO, exe & \ding{55} & \makecell[c]{Honest Server\\Semi-honest Client} & \ding{55} & {\fontsize{4.5}{5.5} \selectfont \makecell[c]{Dataset: CIFAR10/100, ImageNet\\Models: InceptionResNetV2, VGG19,\\MobileNetV2, NASNet\\Platflorm: Intel SGX, TensowFlow\\Baseline: DarkneTZ}} & 2.2-7.8x speedup \\ \hdashline[0.1pt/0.6pt]
    {[SecureFL \cite{kuznetsov2021securefl}, SEC, 2021]}                & SDT & \makecell[c]{Training\\Inference} & TEE & AES & IO, exe & IO, exe & \ding{55} & \makecell[c]{Semi-honest\\Server/Client} & \ding{55} & {\fontsize{4.5}{5.5} \selectfont \makecell[c]{Dataset: MNIST, CIFAR10, Tiny ImageNet\\Models: LeNet, VGG7/16\\Platform: Intel SGX, Raspberry Pi 3B+\\Baseline: Plaintext Model}} & 0.1-0.3x slowdown \\ \hdashline[0.1pt/0.6pt]
    {[Darknight \cite{hashemi2021darknight}, MICRO, 2021]}              & SOT, SOI & \makecell[c]{Training\\Inference} & TEE & MM & IO, exe & IO, exe & \ding{55} & Malicious Server & \ding{51} & {\fontsize{4.5}{5.5} \selectfont \makecell[c]{Dataset: CIFAR10, ImageNet\\Models: VGG16, ResNet50, MobileNetV2\\Platform: Intel SGX, GTX1080 Ti\\Baseline: Fully in Enclave}} & {\fontsize{5.5}{5.5} \selectfont \makecell[c]{Training: 6.5x speedup\\Inference: 12.5x speedup}} \\ \hdashline[0.1pt/0.6pt]
    {[Plinius \cite{yuhala2021plinius}, DSN, 2021]}                     & SOT, SOI & \makecell[c]{Training\\Inference} & TEE & \makecell[c]{AES-GCM,\\PM} & IO, exe & IO, exe & \ding{55} & Malicious Server & \ding{55} & {\fontsize{5.5}{6.5} \selectfont \makecell[c]{Dataset: MNIST\\Models: CNN\\Platform: Intel SGX SDK\\Baseline: Fully in Enclave}} & 2.5-3.5x Speedup \\ \hdashline[0.1pt/0.6pt]
    {[Goten \cite{ng2021goten}, AAAI, 2021]}                            & SOT, SOI & \makecell[c]{Training\\Inference} & TEE & ASS & IO, exe & IO, exe & \ding{55} & Semi-honest server & \ding{55} & {\fontsize{4.5}{5.5} \selectfont \makecell[c]{Dataset: CIFAR10 \\ Models: VGG11 \\ Baseline: CaffeScone, Falcon}} & 6.84-132.64x speedup \\ \bottomrule
\end{tabular*}

\begin{tablenotes}[para]
    \scriptsize
    \item \textbf{ECDH}: Elliptic Curve Diffie-Hellman, \textbf{ZSM}: Zero-sum Masking, \textbf{TSH}: Tree-structured Hierachical, \textbf{ASS}: Additive Secret Sharing, \textbf{MM}: Matrix Masking;
  \end{tablenotes}
      \end{threeparttable}
      \end{table}
    \end{landscape}

    \newpage
    
  \begin{landscape}
    \begin{table}[t]
      \begin{threeparttable}
        \setlength\tabcolsep{2.5pt}
\renewcommand{\arraystretch}{1.3}
\tiny
\begin{tabular*}{\linewidth}{cccccccccccc}
  \toprule
  \multirow{3}*{\scriptsize Schemes} & \multicolumn{2}{c}{\multirow{2}*{\scriptsize Settings}} & \multicolumn{2}{c}{\multirow{2}*{\scriptsize Methodology}} & \multicolumn{5}{c}{\scriptsize Security} &  \multicolumn{2}{c}{\multirow{2}*{\scriptsize Performance}} \\ \cmidrule(lr){6-10} 
        &   &   &   &   & \multicolumn{3}{c}{Security Goals} & \multirow{2}*{\makecell[c]{Adversarial\\Models}} & \multirow{2}*{\makecell[c]{Formal\\proof}} &   & \\ \cmidrule(lr){2-3} \cmidrule(lr){4-5} \cmidrule(lr){6-8} \cmidrule(lr){11-12}  
        & Framework & Function & Technologies & Primitives & Privacy & Integrity & Freshness & & & Configurations & Costs \\ \midrule
    {[PPFL \cite{mo2021ppfl}, MobiSys, 2021]}                           & SDT & Training & TEE & AES-GCM & IO, exe & IO, exe & \ding{55} & \makecell[c]{Semi-honest\\Server/Client} & \ding{55} & {\fontsize{5.5}{6.5} \selectfont \makecell[c]{Dataset: MNIST, CIFAR10\\Models: LeNet, AlexNet, VGG19\\Platform: ARM TrustZone\\Baseline: Fully in Enclave}} & 1.5-3x slowdown \\ \hdashline[0.1pt/0.6pt]
    {[\textsc{Perun} \cite{ozga2021perun}, DBSec, 2021]}                & SOT, SOI & \makecell[c]{Training\\Inference} & TEE & - & IO, exe & IO, exe & \ding{55} & Semi-honest & \ding{55} & {\fontsize{5.5}{6.5} \selectfont \makecell[c]{Dataset: CIFAR10, Medical dataset\\Models: 2D-U-Net\\Platform: TensowFlow, Intel SGX}} & {\fontsize{5.5}{6.5} \selectfont \makecell[c]{Plaintext Model: 0.96x slowdown \\secureTF : 1560x}} \\ \hdashline[0.1pt/0.6pt]
    {[\textsc{Soteria} \cite{brito2021soteria}, ePrint, 2021]}          & SDT & Training & TEE & AES-GCM & IO, exe & IO, exe & \ding{55} & Malicious Server & \ding{55} & {\fontsize{5.5}{6.5} \selectfont \makecell[c]{Dataset: no given\\Models: ALS, PCA, GBT, LR, Naïve Bayes, LDA, K-means\\Platform: Intel SGX SDK, Aparch spark\\Baseline: Fully in Enclave}} & 1.7x speedup \\ \hdashline[0.1pt/0.6pt]
    {[FAML \cite{park2022fairness}, WWW, 2022]}                         & SOI & Inference & TEE & - & IO & IO & \ding{55} & Malicious Server & \ding{55} & {\fontsize{5.5}{6.5} \selectfont \makecell[c]{Dataset: Adult, Bank, COMPAS, German, LSAC\\Models: LR, SVM, NN\\Baseline: Plaintext Model}} & 1.18-2.43x slowdown \\ \hdashline[0.1pt/0.6pt]
    {[AsymML \cite{niu20223lerace}, PETS, 2022]}                        & SOT & Training & TEE & DP & IO, exe & IO, exe & \ding{55} & Malicious Server & \ding{51} & {\fontsize{5.5}{6.5} \selectfont \makecell[c]{Dataset: CIFAR10, ImageNet\\Models: VGG16/19, ResNet18/34\\Platforms: Intel SGX, Pytorch \\Baseline: Fully in Enclave}} & 5.8-7.6x speedup \\ \hdashline[0.1pt/0.6pt]
    {[GINN, \cite{asvadishirehjini2022ginn}, CODASPY, 2022]}            & SOT & Training & TEE & \makecell[c]{SKE/PKE\\SHA256} & IO, exe & IO, exe & \ding{55} & Malicious Server & \ding{51} & {\fontsize{5.5}{6.5} \selectfont \makecell[c]{Dataset: MNIST, CIFAR10\\Models: VGG16/19, ResNet152, ResNet34s\\Baseline: Fully in Enclave}} & 2x-20x speedup \\ \bottomrule
  \end{tabular*}

\begin{tablenotes}[para]
  \scriptsize
  \item \textbf{PM}: Persistent Memory, \textbf{DP}: Differential Privacy, \textbf{SKE}: Symmetric Encryption, \textbf{PKE}: Public Key Encryption;
\end{tablenotes}
      \end{threeparttable}
      \end{table}
    \end{landscape}

Besides, CryptFlow \cite{kumar2020cryptflow} leverages secret sharing to distribute the sensitive input data to several servers and then achieves correctness and confidentiality by cryptographic primitives and integrity by TEE technologies. CryptFlow views the inference as one iteration of training, therefore their method is also suitable for secure multiparty training.

\subsection{Comparisons}

A comprehensive comparison is made in Table \ref{tbl:comp_ppml} to compare existing secure computation protocols for machine learning tasks (including training and inference) with regard to our assessment criteria. 
It is not hard to employ TEE to achieve basic confidentiality and integrity, thus most of the protocols are designed against malicious parties. However, the references against side-channel attacks and replay attacks are very scarce, except for \cite{ohrimenko2016oblivious,hunt2020telekine}. Besides, the references \cite{gu2018securing,florian2019slalom,sun2020shadownet,hashemi2021darknight,niu20223lerace,asvadishirehjini2022ginn} also present formal proof security analysis for their proposals. As for performance, 
we observe that most of the references compare their protocol with either the plaintext or the fully in enclave baseline. 

\section{Conclusions and Future Directions} \label{sec:conclusion}

The Trusted Execution Environment (TEE) has become increasingly popular for both academic research and industrial adoption to achieve secure computation with confidentiality and integrity guarantees. This survey provided a comprehensive overview of state-of-the-art TEE-enabled secure computation protocols. Rest on the proposed taxonomy and systematical assessment criteria, we conducted comprehensive comparisons on TEE-enabled secure computation for both general-purpose function and specific ones, including secure machine learning and secure database queries which have wide applications in practice. 
With the maturing and further development of TEE technologies, we envisage that TEE-enabled secure computation would become a mainstream for practical adoption in the near future. Below we discuss several remaining challenges and future research directions of this promising technology.

\begin{itemize}
\item \textit{General-purpose computation on large-scale data.} As shown in the survey, existing solutions for general-purpose secure computation are only runnable on small-scale data. 
However, it is still very challenging to design general-purpose protocols for large-scale data sets.
Although moderately weakening the security level to gain  performance improvement is a viable strategy, balancing the trade-off between security and performance is known non-trivial task for a long time \cite{deutch2021optimizing}. 
\item \textit{Incremental computation for real-time systems.} Since data is growing daily, it is more practical to perform secure computation based on the previously computed results and the new data. Incremental computation \cite{naor2019incrementally} takes the new data and the previous results as inputs to derive the updated results. We observe that many existing real-time systems employ the incremental algorithm to achieve real-time response. Employing TEE to design secure incremental computation protocols is very interesting and promising for securing real-time applications. Besides, the technique could also be applied to privacy preserving continual learning and reinforcement learning.
\item \textit{Mobile-friendly computation.} Globally, mobile devices are significantly outnumbering desktops and laptops. Although the computing power of mobile devices have kept being improved, they still have limitations in terms of capacity and resource, such as limited power supply and limited space for cooling, etc. Utilizing TEE in mobile computing \cite{chakraborty2019simtpm,atamli2016analysis} and designing a mobile-firendly secure computation framework (e.g., offloading \cite{kumar2013survey}) has not been well investigated and is worth exploring.
\item \textit{System protection against untrusted computation.} Secure computation assumes that TEE's host is malicious and thus achieves confidential computing on sensitive data. In contrast, the untrusted computation means that the program loaded into the TEE is destructive (e.g. Trojan Horse) and thus steals sensitive information from the host. Designing novel approaches to prevent TEE-based untrusted computation is another topic that demands further investigation \cite{hunt2016ryoan, park2020nested}.
\item \textit{TEE trust management.} TEE provides the confidentiality and integrity guarantee of secure computation. However, in practice, not all parties fully trust the TEE. For instance, parties may worry that the TEE is comprised/controlled by another party and thus have no trust on TEE \cite{wu2022hybrid}.  On the other hand, the minimal trust on TEE proposed in \cite{kumar2020cryptflow}  means that secure computation protocols only make trust assumptions on the integrity protection of the hardware, and the confidential computing is shifted to the protocol design. Weakening or diversifying the trust on TEE is a promising direction for future research.
\end{itemize}

\begin{acks}
This reserch is supported by the AXA Research Fund and the National Natural Science Foundation of China, No. 62202358.
\end{acks}

\bibliographystyle{ACM-Reference-Format}
\bibliography{sample-base}

\appendix

\section{Secure Database Queries using TEEs} \label{sec:database}

In this section, we review the TEE-based secure computation for handling encrypted database queries.
Unlike general-purpose secure computation, secure database query protocols aim to implement \textit{data query tasks} with optimal performance.
Over the past two decades, the notion of encrypted databases \cite{popa2011cryptdb} has been proposed to enhance data availability while keeping data confidential. It not only provides a hardware and software infrastructure for securely storing sensitive data, such as personal records, financial documents, and government data, but also offers data query services to authorized users.
As before, we divide secure database queries into three categories: secure outsourced queries, secure distributed queries, and secure multiparty queries.

\subsection{Secure Outsourced Queries}

\subsubsection{Problem Statement}
Generally, a Secure Outsourced Queries (SOQ) protocol consists of three entities: a data owner, a client, and a cloud server. First, the data owner encrypts the original database $\bm{d}$ and then uploads its encrypted form $(\enc{\bm{d}}, \enc{\bm{\Lambda}})$ to the cloud server, in which $\enc{\bm{d}}$ is the encryption of original data and $\enc{\bm{\Lambda}}$ is an encrypted index allowing server to provide query services efficiently. Once the database is uploaded, any authorized client can submit an encrypted query $\enc{q}$ to the cloud server. The server performs the query services in encrypted form and obtains $(\enc{y}, \Gamma) = \Phi\left((\enc{\bm{d}}, \enc{\bm{\Lambda}}), \enc{q}\right)$, where $\Phi$ is a well-designed secure outsourced analytics protocol and $y$ is the query result. Obliviously, it is a special case of secure outsourced computation.
The {\itshape correctness} means that the accurate query results are returned to the client, and the {\itshape completeness} means that all correct results should be returned.
The {\itshape privacy} means that the cloud server cannot learn anything sensitive information about $q$ (query privacy), $(\bm{d}, \bm{\Lambda})$ (database privacy), and $y$ (output privacy). 
The {\itshape integrity} forces the cloud server to run the program $\Phi$ honestly, where $\Gamma$ is the proof for proving the integrity of execution.

The early SOQ protocols \cite{popa2011cryptdb,arasu2015transaction} are based on various cryptographic primitives, for example, searchable encryption, homomorphic encryption, structured encryption, order-preserving encryption for comparison-based queries, accumulator, and Merkle-hash proof for providing integrity guarantee. 
However, they just support some simple queries and fail to answer complex queries efficiently, such as queries involving join operations. Since the emergence of trusted hardware, TEE-based SOQ has provided a promising way to achieve a practical encrypted database while providing privacy protection and integrity guarantee. Besides, TEE is also a natural choice for implementing complex access-control policies in the encrypted database.

\begin{figure}[h]
  \centering
  \includegraphics[width=0.7\linewidth]{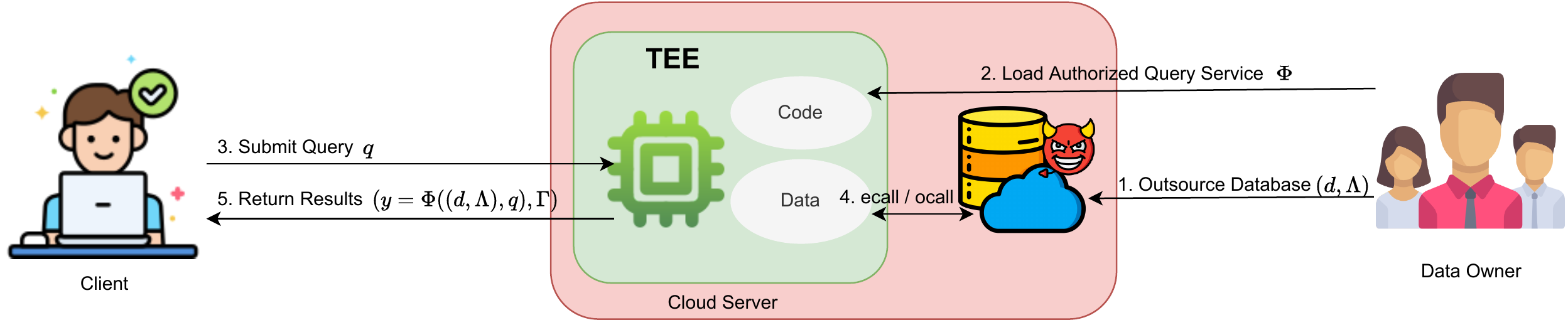}
  \caption{Secure Outsourced Queries}\label{fig:SOQ}
  \Description{Assessment}
\end{figure}

As shown in Fig. \ref{fig:SOQ}, we provide the basic SOQ workflow to explain the interactions between a data owner, a client, and a TEE-equipped server.
(1) The data owner outsources database $(\bm{d}, \bm{\Lambda})$ to the cloud server. Note that the encrypted database is stored on the untrusted domain.
(2) The data owner initializes the TEE and then proves the exact program $\Phi$ for providing SOQ services has been loaded into server's enclave successfully.
(3) The client also builds a secure channel with TEE via remote attestation and then sends a query $q$ to the cloud. (4) The cloud server employs TEE to execute the SOQ protocol honestly, which is run between the EPC and unprotected memory by ecall / ocall invocations. (5) Finally, the enclave returns the final results $(y, \Gamma)$, in which $y = \Phi((\bm{d}, \bm{\Lambda}), q)$ and $\Gamma$ proves the program is executed honestly. Note that all messages communicated with the enclave are sent over the secure channel.

\subsubsection{Research Status}
~\\
\textbf{Relational Database.} In \cite{priebe2018enclavedb}, EnclaveDB ensures the data and the query are confidential, intact, and fresh, even if the host is malicious. The database logs are verified for achieving integrity and freshness. EnclaveDB presents a solution to implement minimal thread synchronization required for concurrent, asynchronous appending and truncation. By experiments, EnclaveDB introduces about 40\% overhead compared to the traditional database systems. However, EnclaveDB only works on the encryption mode, and the access pattern may leak sensitive information to the attackers.

To mitigate the side channel attacks, ObliDB was presented to provide efficient and secure SQL queries via multiple access methods \cite{eskandarian2019oblidb}. For example, four oblivious SELECT algorithms are embedded in the database system, obliDB takes advantage of knowledge about the query to choose an optimal algorithm to finish the SELECT operation. Therefore the methods apply to different settings. However, the structure of the queries is leaked to the adversary. In \cite{dang2017privacy}, Dang et al. presented an optimal computational model called Scramble-then-Compute (STC). STC introduced a new component (called Scrambler) to prevent leakage from access patterns, and the main idea is to permute the input data randomly before the original algorithm is invoked. In such a way, the access pattern does not leak any sensitive information while the TCB is kept lean. With STC, one can improve the efficiency of an extensive body of oblivious algorithms (e. g. sort, compaction, selection, aggregation, join). In \cite{vinayagamurthy2019stealthdb}, a scalable encrypted database (StealthDB) was presented to support full SQL queries, which include query services as well as database creation and transaction processing. While ObliDB achieves the strongest protection by making the user data completely inaccessible, StealthDB offers add-on features but only achieves weaker protection. In \cite{sun2021building}, Sun et al. explored the trade-offs in security, performance, and functionality. They proposed the Enclage, which leverages B+ tree to design a three-tier index (record level, buffer level, and page level). With the three-tier index, they studied the optimal node size, buffer size, and page size. Extensive experiments showed that Enclage achieves 5-9x improvements than previous schemes.

Besides, VeriDB \cite{zhou2021veridb} studies the verifiable general queries technologies on the relational databases via SGX. VeriDB ensures the correctness, completeness, and freshness of the query results. Note that VeriDB focuses on the verifiability of queries and is built on top of an unencrypted database. Thus it fails to provide the privacy proetction of the database. To achieve verifiability, VeriDB only incurs only about 1-2 ms on read/write operations and 9\%-39\% overhead for the complex analytical workload.

\noindent\textbf{Key-Value Store.} To ensure confidentiality and integrity of key-value store management, a basic strategy is to load the entire key-value store to the enclave. Then a client can read or write the data via the enclave interfaces. However, the performance is very low when a large store is maintained because of the limited EPC and the page switches. ShieldStore \cite{kim2019shieldstore} manages the key-value stores by the hash table, and it overcomes the EPC limits by maintaining an encrypted hash table in untrusted memory associated with Merkle trees for integrity protection. ShieldStore only stores the header pointer of the hash table and Merkle tree root nodes in the enclave. With the protection of enclaves, ShieldStore outperforms 8-11 times than baseline SGX key-value stores. In order to ensure freshness against rewind attacks, SPEICHER \cite{bailleu2019speicher} borrows the idea from RocksDB by embedding an asynchronous trusted counter to an authenticated and confidentiality-preserving LSM data structure, which is maintained in the untrusted storage medium. Compared with the RocksDB, SPEICHER only incurs reasonable overheads for ensuring confidentiality, integrity, and freshness.

\noindent\textbf{Labeled Documents.} Labeled documents consist of many documents, and each is associated with one or several keyword(s). Inverted index design is very popular in labeled document retrieval systems. An inverted index contains a set of keywords, and a document posting list of each keyword represents documents that contain the keyword. Shao et al. \cite{shao2020index} proposed the masked inverted index (MII), which supports efficient query processing with the help of SGX.

In \cite{ferreira2020boolean}, Ferreira et al. designed the BISEN, a new Boolean Symmetric Searchable Encryption (SSE) scheme based on trusted hardware and the traditional SSE. BISEN supports multiple clients with access control features, provides verifiability against fully malicious adversaries, supports dynamic updates with forward and backward privacy, and supports arbitrarily complex boolean queries with filters. Note that BISEN only reveals which encrypted index entries are accessed. By leveraging TEEs as remote trust anchors, BISEN is able to move most client-side computations to the server, reducing computation, storage, and communication overheads. {\itshape In fact, BISEN builds the index on cloud server}. Besides, \cite{fuhry2017hardidx} and \cite{kim2019shieldstore} showed how to build B+ tree and TF-IDF securely using Intel SGX in remote servers.

\subsection{Secure Distributed Queries}

\subsubsection{Problem Statement}
Secure distributed queries (SDQ) is run between a master node and multiple slave nodes, similar as SDC described in Sec. 3.2. The data owner outsources the encrypted database to the master node and authorities the query services to the master node. The master node splits the large-scale database into small-scale databases and then distributes them to the slave nodes.
When receiving a query $q$ from a client, each slave node computes an intermediate result based on its inputs. Then the master node aggregates all intermediates into the final results. Precisely, an SDQ protocol consists of two sub-protocols $\Psi$ and $\Pi$, where $\Psi$ is run in slave nodes for obtaining the intermediate results, and $\Pi$ is run between the master node and the slave nodes for getting the final results. Specifically,
\begin{equation}
\Phi\left((\bm{d}, \bm{\Lambda)}, q\right) = \Pi\left(\Psi((\bm{d_1}, \bm{\Lambda_1}), q),\Psi((\bm{d_2}, \bm{\Lambda_2}), q), \cdots, \Psi((\bm{d_N}, \bm{\Lambda_N}), q)\right)
\end{equation}
Where $(\bm{d}, \bm{\Lambda}) \triangleq (\bm{d}_1, \bm{\Lambda}_1) \circ \cdots \circ  (\bm{d}_N, \bm{\Lambda}_N)$ \footnote{$\circ$ denotes the joint operation between the databases.}.
Similar to SDC, the privacy and integrity also should be achieved in an SDQ protocol.

\begin{figure}[h]
  \centering
  \includegraphics[width=\linewidth]{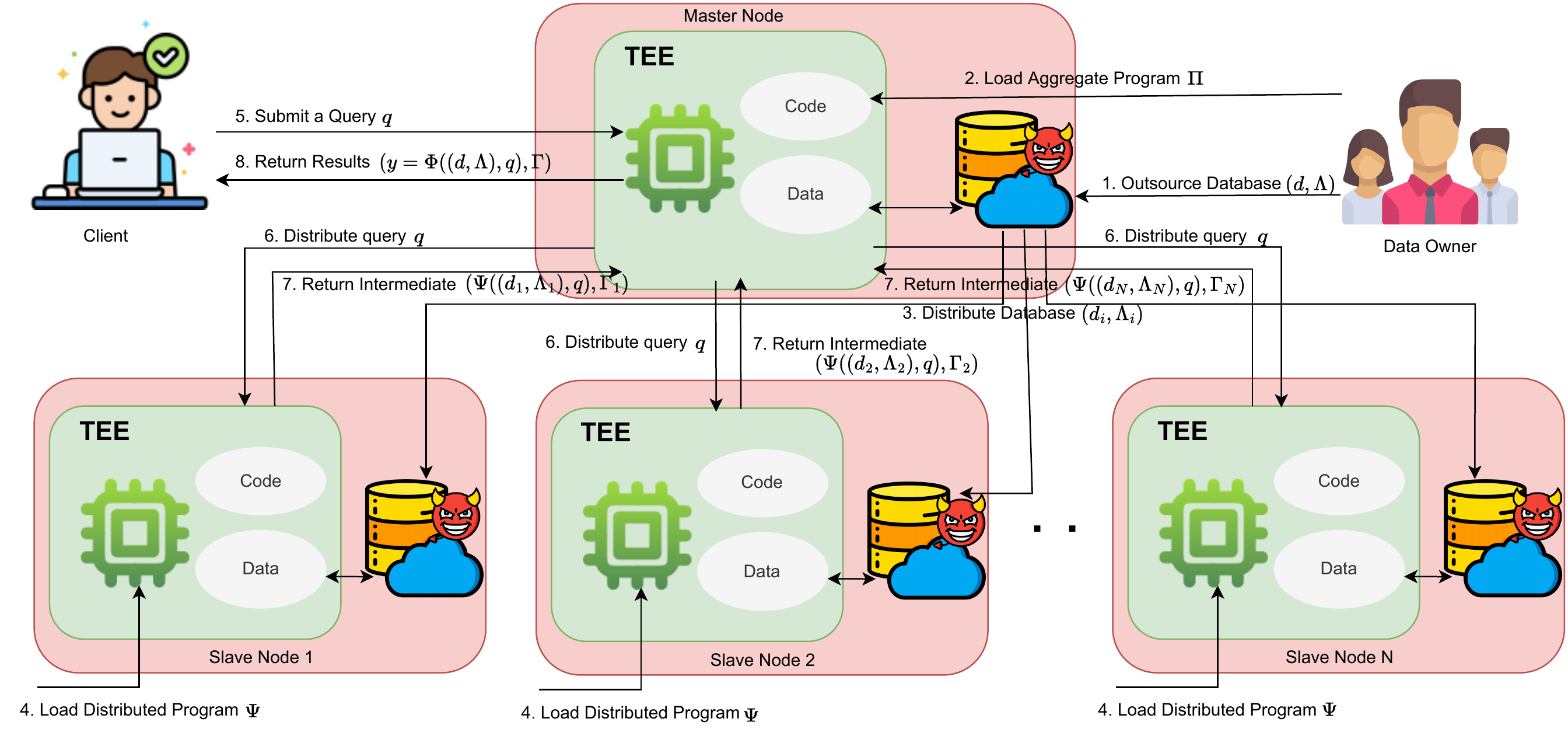}
  \caption{Secure Distributed Queries}\label{fig:SDQ}
  \Description{Assessment}
\end{figure}

While many primitives to design SDQ protocols are similar to those for SOQ in an outsourced framework, SDQ protocols focus more on the compatibility with some popular distributed platforms such as MapReduce and Spark. Therefore, besides methods for achieving secure queries, we also care about whether a protocol is compatible with existing distributed platforms.
As shown in Fig. \ref{fig:SDQ}, we provide the basic SDQ workflow to explain the interactions among a data owner, a client and the TEE-equipped nodes.
(1) The data owner outsources database $(\bm{d}, \bm{\Lambda})$ to master node. Note that the encrypted database is stored on the untrusted domain.
(2) The data owner initializes the TEE and then proves the exact program $\Pi$ for providing aggregation SDQ services has been loaded into master's enclave successfully.
(3) The master node splits the database $(\bm{d}, \bm{\Lambda})$ into small-scale databases $(\bm{d}_i, \bm{\Lambda}_i)$, $i = 1, 2, \cdots, N$.
(4) The master node also builds secure channel via remote (or local) attestation with the slaves and then ensures the distributed program $\Psi$ is loaded into slaves' TEE successfully.
(5) The client also builds a secure channel with TEE via remote attestation and then sends a query $q$ to master node.
(6) Then The master node sends $q$ to all slave nodes, and each slave node obtains an intermediate result $(\Psi\left((\bm{d}_i, \bm{\Lambda}_i), q\right), \Gamma_i)$, which is run between the enclave and untrusted host by ecall/ocall invocations. Note that $\Gamma_i$ is employed to prove the program is run honestly in each slave node.
(7) After that, $\Pi$ is called to aggregate all intermediate results into the final results.
(8) Finally, the master node returns the final results $(y, \Gamma)$, in which $y = \Phi((\bm{d}, \bm{\Lambda}), q)$ and $\Gamma$ proves the program is executed honestly.
Noted all messages communicated with enclave are sent over the secure channel.

\subsubsection{Research Status}
In \cite{zheng2017opaque}, Zheng et al. proposed the distributed analytics platform Opaque. Opaque promises obliviousness of SQL operations and thus achieves a strong privacy guarantee. To ensure obliviousness, Opaque introduces the Intra-machine oblivious sorting algorithm and the Inter-machine oblivious sorting algorithm, which serve as the basic building blocks for the oblivious filter, oblivious aggregation, and oblivious join. Furthermore, they also designed novel query planning techniques that improve efficiency by reducing the number of sort invocations and introducing mixed sensitive attributes. The experimental results showed that obliviousness by fully-padding mechanism comes with a 1.6-46x overhead. The fully-padding mechanism prevents information leakage from the size of intermediate results. To further improve the efficiency, Hermetic \cite{xu2019hermetic} applied the differential private (DP) padding mechanism to the distributed data analytics. The DP-padding method achieves 1.6-43x improvement than the fully-padding mechanism. When Hermetic only applies DP to pad the intermediate results, DPSpark \cite{wu2021differentially} also applies it to the algorithms and thus introduces the notion of ($\epsilon, \delta$)-differentially private obliviousness (($\epsilon, \delta$)-DPO), which relaxes full obliviousness to enable efficiency improvements. They first presented the perturbation-shuffle-analysis framework and then designed several differentially oblivious operators (such as sort and join). The DPSpark system is optimized by reducing the number of oblivious shuffles and is benchmarked in different parameters. In \cite{allen2019algorithmic}, several classical algorithms were implemented with differential obliviousness.

For Key-value stores, Bailleu et al. proposed Avocado \cite{bailleu2021avocado}, which overcomes the physical memory limitation and thus achieves high-performance  and fault-tolerance in a malicious environment. By experiments, Avocado provides a fast lookup speed of about 1.5x-9x faster than ShieldStore. For stream data, Part et al. presented the StreamBox-TZ \cite{park2019streambox}, which processes large stream data in an IoT environment. StreamBox-TZ solves two major challenges: minimizes the TCB in edge devices and verifies the execution of stream analytics on edge. Moreover, on the octa-core ARMv8 platform, StreamBox-TZ exhibits performance with sub-second delay.

\subsection{Secure Multiparty Queries}

\subsubsection{Problem Statement}

Like SMC, a secure multiparty queries (SMQ) protocol also targets answering a joint database query securely among several distinct yet connected parties. SMQ is usually designed to handle complex queries. Specifically, each participant has a local database $(\bm{d}_i, \bm{\Lambda}_i)$ and the goal is to design a protocol to answer a query $q$ on the joint database $(\bm{d}, \bm{\Lambda}) \triangleq (\bm{d}_1, \bm{\Lambda}_1) \circ \cdots \circ  (\bm{d}_N, \bm{\Lambda}_N)$. 
The correctness means that the SMQ protocol completes the analytical tasks and obtains the final result correctly, and privacy means that the execution of the protocol only reveals the final result to the intended parties. Integrity ensures that all parties follow the protocol steps honestly. Finally, fairness is also defined in SMQ, which means malicious parties receive the outputs if and only if other honest parties also receive their outputs.

\begin{figure}[h]
  \centering
  \includegraphics[width=0.7\linewidth]{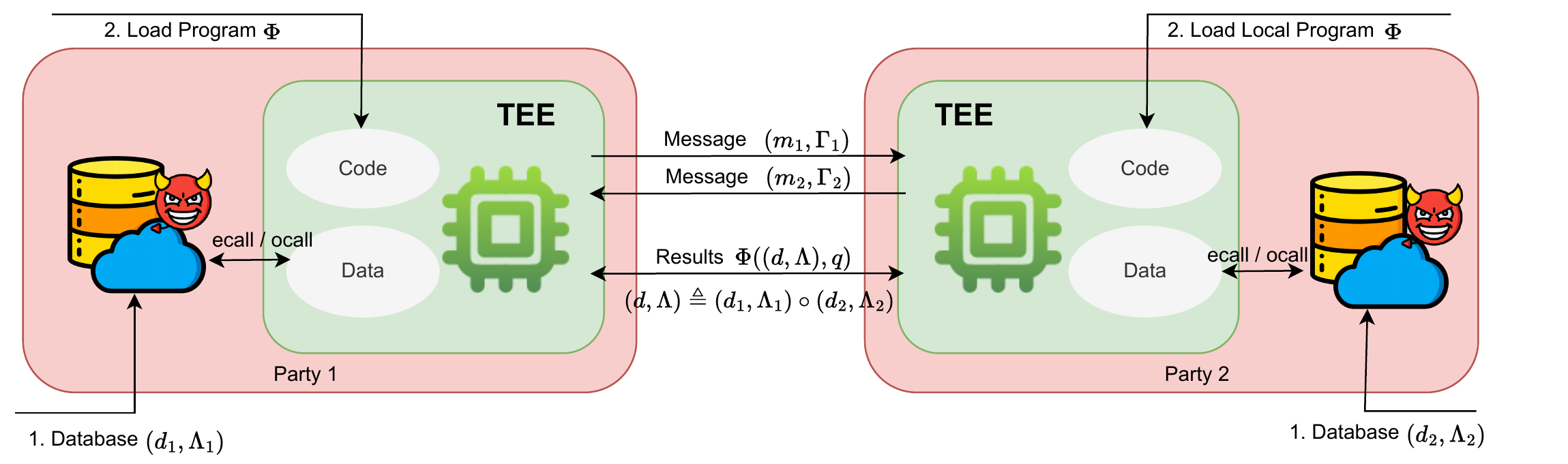}
  \caption{Secure Multiparty Queries}\label{fig:SMQ}
  \Description{Assessment}
\end{figure}

Many cryptographic SMC protocols can be leveraged to design SMQ protocols \cite{bater2017smcql,volgushev2019conclave}; however, they are still impractical because of either the huge communication cost or the extensive computational cost. In this survey, we focus on the hardware-assisted approaches, which improve efficiency significantly. Fig. \ref{fig:SMQ} shows the workflow of a secure two-party SMQ protocol. 
The two parties each has a database, denoted as $(\bm{d}_1, \bm{\Lambda}_1)$ and $(\bm{d}_2, \bm{\Lambda}_2)$, respectively. Then the local program $\Phi$ is loaded into the enclave. When receiving a query $q$, the two parties finish the computation in an interactive manner, which consists of multiple rounds of communication.
In each round, a message-proof tuple $(m_i, \Gamma_i)$ is derived from its private inputs, and then the tuple is sent to another party. $\Gamma_i$ is employed to prove the message $m_i$ is computed honestly. Finally, the two parties may run a fair protocol to distribute the final result $\Phi((\bm{d}, \bm{\Lambda}), q)$, where $(\bm{d}, \bm{\Lambda}) \triangleq (\bm{d}_1, \bm{\Lambda}_1) \circ  (\bm{d}_2, \bm{\Lambda}_2)$ is the joint database.
Noted that all messages between the two parties are sent over the secure channel.

\subsubsection{Research Status} While protocols (such as SFA \cite{rishabhpoddarsecure}, Shrinkwrap \cite{bater2018shrinkwrap}, ObliJoin \cite{krastnikov2021efficient,chang2022towards}) are based on differential privacy or cryptographic methods, OCQ proposed in \cite{dave2020oblivious} leveraged hardware enclaves to build a general framework for secure multiparty analytics. Since SMQ is executed among different parties and thus it always results in high communication costs. In OCQ, parties involved in the protocol should share a schema in advance, then execute the well-designed protocols obliviously for some allowed queries in the wide-area network. They also presented an optimized query planner with sensitive cardinalities, which executes the queries based on relational operators in order to prevent data leakage caused by side channels such as memory access patterns, network traffic, and other factors such as cardinality and statistics. Compared to the Opaque, OCQ is up to 9.9x faster  in the wide-area network. However, the improvement is only applicable for selection and aggregation queries, and for the join queries, the cost is still not practical. The works discussed in Sec. 3.3 are workable for data analytics. However, they only work for very small-scale data sets. Although some SDQ protocols can be extended to SMQ protocols, designing practical SMQ protocols is still very challenging. Generally, it remains an open problem to design practical SMQ protocols with high throughput.

\subsection{Comparisons}

We present a comprehensive comparison of secure database queries protocols based on their setting, methodology, security features, and performance. Table \ref{tbl:comp_sa} demonstrates that many of current hardware-supported encrypted database systems focus on the SQL queries over relation databases, and some of them focus on set/get operations over key-value stores or keyword queries over labeled documents set. However, there are very few studies on analyzing the steam data, except for \cite{park2019streambox}. From the security perspective, keeping input/output privacy is the basic security requirement of TEE-based protocols. Employing oblivious primitives provides stronger privacy protection, for example some SOQ protocols (e.g. ObliDB \cite{eskandarian2019oblidb}, Enclage \cite{eskandarian2019oblidb}) can avoid access pattern leakage in the memory level, whereas some SDQ (e.g. Hermetic \cite{xu2019hermetic}, DPSpark \cite{xu2019hermetic}) and SMQ (e.g. OCQ \cite{dave2020oblivious}) protocols prevent pattern leakage from both memory and network level. While most of the existing protocols achieve integrity of IO and execution, EnclaveDB \cite{priebe2018enclavedb} and VeriDB \cite{zhou2021veridb} also achieve the query integrity. From performance aspects, we summarize the protocol costs with regards to different configurations and different functions. If a protocol is compared to the baseline ``fully in enclave'', we employ a positive number to represent their improvements. For example, Enclage \cite{eskandarian2019oblidb} improves the baseline about 5-13 times. If a protocol is compared to the baseline ``plaintext model'', we employ a negative number to represent their overhead. For example, Opaque \cite{priebe2018enclavedb} introduces about -160\% -- -460\% overhead compared to the baseline on join operations.

\begin{landscape}
\begin{table}[t]
  \caption{Comparisons of Secure Data Queries}
  \begin{threeparttable}
    \renewcommand{\arraystretch}{1.3}
\setlength\tabcolsep{1.5pt}
\tiny
\begin{tabular*}{\linewidth}{cccccccccccc}
  \toprule
  \multirow{3}*{\scriptsize Schemes} & \multicolumn{2}{c}{\multirow{2}*{\scriptsize Settings}} & \multicolumn{2}{c}{\multirow{2}*{\scriptsize Methodology}} & \multicolumn{5}{c}{\scriptsize Security} &  \multicolumn{2}{c}{\multirow{2}*{\scriptsize Performance}} \\ \cmidrule(lr){6-10} 
        &   &   &   &   & \multicolumn{3}{c}{Security Goals} & \multirow{2}*{\makecell[c]{Adversarial\\Models}} & \multirow{2}*{\makecell[c]{Formal\\proof}} &   & \\ \cmidrule(lr){2-3} \cmidrule(lr){4-5} \cmidrule(lr){6-8} \cmidrule(lr){11-12}  
        & Framework & Function & Technologies & Primitives & Privacy & Integrity & Freshness & & & Configurations & Costs \\ \midrule
    {[Opaque \cite{zheng2017opaque}, NSDI, 2017]}                     & SDA, n & SQL & Stateful TEE + OP & AE, oSort & \makecell[c]{IO, mPattern,\\nPattern, IR} & IO, exe & \ding{51} & Malicious Server & \ding{55} & {\fontsize{4.8}{5.3}  \selectfont \makecell[c]{DataSet: Relational Database\\Platform: Spark\\Number of nodes: 5\\Baseline: Plaintext Model}} & {\fontsize{4.8}{5.3} \selectfont \makecell[c]{Join: 1.6-4.6x slowdown \\PageRank: 3.8x slowdown}} \\ \hdashline[0.1pt/0.6pt]
    {[STC \cite{dang2017privacy}, PETS, 2017]}                        & SOA & SQL & Stateful TEE + OP & MS, PRP & IO, mPattern & IO, exe & \ding{55} & \makecell[c]{Semi-honest,\\Adaptive Server} & SIM & {\fontsize{4.8}{5.3} \selectfont \makecell[c]{DataSet: Relational Database\\Platform: Spark\\Baseline: Plaintext Model}} & {\fontsize{4.8}{5.3} \selectfont \makecell[c]{Sort: -79\%\\Compaction: 3.9x slowdown\\Select: 2.4x slowdown\\Aggregation: 1.2x slowdown\\Join: 3.8x slowdown}}\\ \hdashline[0.1pt/0.6pt]
    {[EnclaveDB \cite{priebe2018enclavedb}, S\&P, 2018]}              & SOA & SQL & Stateful TEE & AEAD & IO, Query & \makecell[c]{IO, exe,\\Log, Query} & \ding{51} & Malicious Server & \ding{55} & {\fontsize{4.8}{5.3} \selectfont \makecell[c]{DataSet: Relational Database\\Platform: Hekaton\\Baseline: Plaintext Model}} & {\fontsize{4.8}{5.3} \selectfont TATP: 0.2x slowdown} \\ \hdashline[0.1pt/0.6pt]
    {[ObliDB\tnote{1} \cite{eskandarian2019oblidb}, VLDB, 2019]}      & SOA & SQL & Stateful TEE + OP & AE, oSort & \makecell[c]{IO, IR,\\mPattern} & IO, exe & \ding{55} & Malicious Server & \ding{55} & {\fontsize{4.8}{5.3} \selectfont \makecell[c]{DataSet: Relational Database\\Platform: Spark\\Baseline: Opaque}} & {\fontsize{4.8}{5.3} \selectfont Join: 1.4-2.7x speedup} \\ \hdashline[0.1pt/0.6pt]
    {[ShieldStore\tnote{2} \cite{kim2019shieldstore}, EuroSys, 2019]} & SOA & Set/Get & Stateful TEE & AE, MAC, MHT & IO & IO, exe & \ding{55} & Malicious Server & \ding{55} & {\fontsize{4.8}{5.3} \selectfont \makecell[c]{DataSet: Key-value store\\Platform: Ubuntu Server\\Number of Entries: 10M\\Baseline: Fully in Enclave}} & {\fontsize{4.8}{5.3} \selectfont \makecell[c]{RD50\_Z: 1.2x speedup\\RD95\_Z: 1.6x speedup\\RD100\_Z: 1.7x speedu}} \\ \hdashline[0.1pt/0.6pt]
    {[SPEICHER \cite{bailleu2019speicher}, FAST, 2019]}               & SOA & Set/Get & Stateful TEE & Skiplist, LSM & IO, State & IO, exe & \ding{51} & Malicious Server & \ding{55} & {\fontsize{4.8}{5.3} \selectfont \makecell[c]{DataSet: RocksDB\\Platform: Ubuntu Server\\Baseline: Plaintext Model}} & {\fontsize{4.8}{5.3} \selectfont \makecell[c]{RD90\_U: 14.8x slowdown\\RD80\_U: 15.2x slowdown\\RD100\_U: 32x slowdown}} \\ \hdashline[0.1pt/0.6pt]
    {[StealthDB \cite{vinayagamurthy2019stealthdb}, PETS, 2019]}      & SOA & SQL & Stateful TEE & AE & IO & exe & \ding{55} & Semi-honest Server & SIM & {\fontsize{4.8}{5.3} \selectfont \makecell[c]{DataSet: Relational Database\\Baseline: Plaintext Model}} & {\fontsize{4.8}{5.3} \selectfont 0.18-1.2x slowdown} \\ \hdashline[0.1pt/0.6pt]
    {[Hermetic \cite{xu2019hermetic} , Tech. Report, 2019]}           & SDA, n & SQL & Stateful TEE + DOP & DP & \makecell[c]{IO, mPattern,\\nPattern, IR} & IO, exe & \ding{55} & \makecell[c]{Semi-honest,\\Adaptive Server} & DP Analysis & {\fontsize{4.8}{5.3} \selectfont \makecell[c]{DataSet: Relational Database\\Platform: Spark\\ \$$(\epsilon, \delta)$: (5e-3, 1e-5)\\Baseline: Fully-Padding in Enclave}} & {\fontsize{4.8}{5.3} \selectfont \makecell[c]{Select: 2.1-3.2x speedup\\Groupby: 1.8—2.7x speedup\\Join: 4.1—8.3x speedup}} \\ \hdashline[0.1pt/0.6pt]
    {[SBT\tnote{3} \cite{park2019streambox}, USENIX ATC, 2019]}       & SDA, n & Stream Algorithms & Stateless TEE & AE & IO & IO,exe & \ding{55} & \makecell[c]{Malicious\\Edge Devices} & \ding{55} & {\fontsize{4.8}{5.3} \selectfont \makecell[c]{DataSet: Taxi IDs\\Platform: octa core ARMv8 p\\Baseline: Plaintext Model}} & {\fontsize{4.8}{5.3} \selectfont \makecell[c]{Top-k: 0.04-0.12x slowdown\\WinSum: 0.5-0.8x slowdown\\Filter: 0.5-0.9x slowdown}} \\ \hdashline[0.1pt/0.6pt]
    {[BISEN\tnote{4} \cite{ferreira2020boolean}, TDSC, 2020]}         & SOA & Boolean Queries & Stateful TEE & AE + PRF & IO, Query & IO, exe & \ding{55} & Malicious Server & \ding{55} & {\fontsize{4.8}{5.3} \selectfont \makecell[c]{DataSet: Labeled Document\\Platform: Ubuntu Server\\Number of documents: 5.5M\\ Baseline: No baseline}} & {\fontsize{4.8}{5.3} \selectfont \makecell[c]{Search: 50s\\Update: 2.7h}} \\ \hdashline[0.1pt/0.6pt]
    {[OCQ \cite{dave2020oblivious}, EuroSys, 2020]}                   & SMA & SQL & Stateful TEE + OP & AE, oSort & \makecell[c]{IO, mPattern,\\nPattern, IR} & IO, exe & \ding{55} & Malicious Server & SIM & {\fontsize{4.8}{5.3} \selectfont \makecell[c]{DataSet: Relational Database\\Platform: Spark\\Number of nodes: 5.\\ Baseline:Opaque}} & {\fontsize{4.8}{5.3} \selectfont \makecell[c]{Aggregation: 6.7—9.9x speedup\\Join: 1.0—2.3x speedup}} \\ \hdashline[0.1pt/0.6pt]
    {[DPSpark \cite{wu2021differentially}, TDSC, 2021]}               & SDA, n & SQL & Stateful TEE + DOP & DP & \makecell[c]{IO, mPattern,\\nPattern, IR} & IO, exe & \ding{55} & \makecell[c]{Semi-honest,\\Adaptive Server} & DP Analysis & {\fontsize{4.8}{5.3} \selectfont \makecell[c]{DataSet: Relational Database\\Platform: Spark\\Baseline: Plaintext model}} & {\fontsize{4.8}{5.3} \selectfont \makecell[c]{WordCount: 0.1-0.2x slowdown\\PageRank: 0.35-0.54x slowdown\\Log Query: 0.25-0.4x slowdown\\K-Means: 0.2-0.5x slowdown\\GroupBy: 0.45-0.85x slowdown}} \\ \hdashline[0.1pt/0.6pt]
    {[MII \cite{shao2020index}, CIKM, 2020]}                          & SOA & Top-k & Stateful TEE & oShuffle & IO, Query & IO, exe & \ding{55} & Semi-honest Server & \makecell[c]{Oblivious\\Analysis} & {\fontsize{4.8}{5.3} \selectfont \makecell[c]{DataSet: Relational Database\\Number of documents: 1.3M\\Baseline: \cite{ding2011faster}}} & {\fontsize{4.8}{5.3} \selectfont Top-k: 0.29-1.5x speedup} \\ \hdashline[0.1pt/0.6pt]
    {[Enclage \cite{sun2021building}, VLDB, 2021]}                    & SOA & Set/Get & Stateful TEE + OP & DE, B+ tree & \makecell[c]{IO, IR,\\mPattern} & IO, exe & \ding{55} & Malicious Server & \ding{55} & {\fontsize{4.8}{5.3} \selectfont \makecell[c]{DataSet: Key-value store\\Platform: Ubuntu Server\\Number of Items: 10M\\Baseline: Fully in Enclave}} & {\fontsize{4.8}{5.3} \selectfont YSCB: 5-13x speedup} \\ \hdashline[0.1pt/0.6pt]
    {[VeriDB \cite{zhou2021veridb}, SIGMOD, 2021]}                    & SOA & SQL & Stateful TEE & MHT & \ding{55} & IO, exe, query & \ding{51} & Malicious Server & \ding{55} & {\fontsize{4.8}{5.3} \selectfont \makecell[c]{DataSet: Relational Database\\Platform: Ubuntu Server\\Number of Items: 10M\\Baseline: No baseline}} & {\fontsize{4.8}{5.3} \selectfont \makecell[c]{Select: 20-30s\\Join: 80-90s}} \\ \hdashline[0.1pt/0.6pt]
    {[Avocado\tnote{5} \cite{bailleu2021avocado}, USENIX ATC, 2021]}  & SDA, n & Set/Get & Stateful TEE & BFT & \makecell[c]{IO, IR,\\mPattern} & IO, exe & \ding{51} & Malicious Server & \ding{55} & {\fontsize{4.8}{5.3} \selectfont \makecell[c]{DataSet: Key-value store\\Platform: Ubuntu Server\\Number of Items: 10M\\Baseline: ShieldStore}} & {\fontsize{4.8}{5.3} \selectfont YSCB: 6-24x speedup} \\ \bottomrule
  \end{tabular*}

\begin{tablenotes}[para]
  \scriptsize
  \item \textbf{AE}: Authenticated Encryption, \textbf{MS}: Melbourne Shuffle, \textbf{PRP}: Pseudo-Random Permutation, \textbf{IR}: Intermediate Results, \textbf{SIM}: Simulate-based Proof, \textbf{TATP}: The workload that consists of 40000 transactions, \textbf{MAC}: Message Authentication Code, \textbf{MHT}: Merkle Hash Tree, \textbf{LSM}: Log-Structured Merge Tree, \textbf{PRF}: Pseudo-Random Functions, \textbf{OP}: Oblivious Primitives, \textbf{DOP}: Differential Oblivious Primitives, \textbf{DE}: Delta Encryption
 , \textbf{oSort}: Oblivious Sort, \textbf{oShuffle}: Oblivious Shuffle, \textbf{BFT}: Byzantine Fault Tolerance; \\
  \item[1] https://github.com/SabaEskandarian/ObliDB.git, \item[2] https://github.com/cocoppang/ShieldStore.git, \item[3] http://xsel.rocks/p/streambox, \item[4] https://github.com/bernymac/BISEN.git, \item[5] https://github.com/mbailleu/avocado.git;
\end{tablenotes}
  \end{threeparttable}\label{tbl:comp_sa}
  \end{table}
\end{landscape}


\end{document}